%% file: Main.tex
\documentclass[12pt, draftclsnofoot, onecolumn]{IEEEtran}
\input{packageList.tex}
\input{commandList.tex}
\newtoggle{oneColumn}
\makeatletter
\if@twocolumn
	\togglefalse{oneColumn}
\else
	\toggletrue{oneColumn}
\fi
\makeatother

\iftoggle{oneColumn}
{
	\title{Downlink Performance of Superimposed Pilots in Massive MIMO Systems}
	\author{Karthik Upadhya,~\IEEEmembership{Student~Member,~IEEE},\\ 
		Sergiy A. Vorobyov,~\IEEEmembership{Fellow,~IEEE}, \\
		Mikko Vehkaper\"a,~\IEEEmembership{Member,~IEEE}
		\thanks{K. Upadhya, S. A. Vorobyov, and M. Vehkaper\"a are with the Department of Signal Processing and Acoustics, Aalto University, FI-00076 Aalto, Finland (E-mails: karthik.upadhya@aalto.fi, svor@ieee.org, mikko.vehkapera@aalto.fi).
	     Parts of this paper were presented at IEEE GlobalSIP, Greater Washington, D.C., 2016 and IEEE ICASSP, New Orleans, Louisiana, 2017.}
	}
}
{
	\title{Downlink Performance of Superimposed Pilots in Massive MIMO Systems}
	\author{Karthik Upadhya,~\IEEEmembership{Student~Member,~IEEE}, 
		Sergiy A. Vorobyov,~\IEEEmembership{Senior~Member,~IEEE},\\
		Mikko Vehkapera,~\IEEEmembership{Member,~IEEE}
		\thanks{K. Upadhya, S. A. Vorobyov, and M. Vehkaper\"a are with the Department of Signal Processing and Acoustics, Aalto University, FI-00076 Aalto, Finland (E-mails: karthik.upadhya@aalto.fi, svor@ieee.org, mikko.vehkapera@aalto.fi).}		
	}
}
\begin{document}
\maketitle
\input{texFiles/abstract.tex}
\input{texFiles/introduction.tex}
\input{texFiles/systemModel.tex}
\input{texFiles/pilotContaminationDownlink.tex}
\input{texFiles/staggeredPilot.tex}
\input{texFiles/hybridSystem.tex}
\input{texFiles/numericalSimulations.tex}
\input{texFiles/conclusion.tex}
\input{texFiles/appendixDlSinrCalcFeedback.tex}
\input{texFiles/appendixRhoDRhoPCalc.tex}
\input{texFiles/appendixMseCrlbFeedback.tex}

\bibliographystyle{IEEEtran}
\bibliography{IEEEabrv,paperBibliography}
\end{document}

%% file: packageList.tex
\usepackage{tikz}
\usepackage{amsmath,amssymb}
\usepackage{pgfplots}
\pgfplotsset{compat=1.7}
\usepackage{amsthm}
\usepackage{cite}
\usepackage{hyperref}
\usepackage{tabularx}
\usepackage{array,booktabs}
\usepackage{tablefootnote}
\usepackage{standalone}
\usepackage{cleveref}
\usepackage{enumitem}
\usepackage{graphicx}
\usepackage{etoolbox}
\usepackage{algpseudocode,algorithmicx,algorithm}
\usepackage{xr}
\usepackage{mdframed}

%% file: commandList.tex
\newcommand{\mbf}[1]{\boldsymbol{#1}}
\newcommand{\lrf}[1]{\left\{#1\right\}}
\newcommand{\lrc}[1]{\left(#1\right)}
\newcommand{\lrs}[1]{\left[#1\right]}
\newcommand{\vecmat}[1]{\overline{\mbf{#1}}}
\newcommand{\indicator}[1]{\mbf{1}_{\lrf{#1}}}
\newcommand{\rth}{\text{th}}

\newcommand{\condProb}{\;\middle\vert\;}
\newcommand{\eye}{\mbf{I}}

\newcommand\custFontSize{\fontsize{7pt}{11pt}\selectfont}

\newtheorem{prop}{Proposition}
\makeatletter
\g@addto@macro\th@remark{\thm@headpunct{\textnormal{:}}}
\makeatother
\theoremstyle{remark}
\newtheorem{remark}{Remark}

\newcommand{\Ugtrless}{%
  \mathrel{\kern0pt\mathop{\gtrless}\limits^{\mathcal{U}_\vIdxOne}_{\overline{\mathcal{U}}_\vIdxOne}}%
}

\newcommand{\gtr}{%
  \mathrel{\kern0pt\mathop{<}\limits^{\mathcal{U}_\vIdxOne}}%
}

\newcommand{\coherenceTime}{C}
\newcommand{\ulDuration}{C_u}
\newcommand{\dlDuration}{C_d}

\newcommand{\pilotReuseRatioTp}{r^{\mathrm{RP}}}
\newcommand{\pilotReuseRatioSp}{r^{\mathrm{SP}}}
\newcommand{\pilotReuseRatio}{r}

\newcommand*{\vIdxOne}{j} 
\newcommand*{\vIdxTwo}{m}
\newcommand*{\sIdxOne}{\ell}
\newcommand*{\sIdxTwo}{k}
\newcommand*{\sIdxThree}{n}
\newcommand*{\sIdxFour}{p}

\newcommand{\expectation}{\mathbb{E}}
\newcommand{\sinrUlNonIterative}{\mathrm{SINR}^{\mathrm{SP-ul}}}
\newcommand{\powerBackoff}{\omega}

\newcommand{\cpUserSet}{\mathcal{U}_{\mathrm{RP}}}
\newcommand{\spUserSet}{\mathcal{U}_{\mathrm{SP}}}
\newcommand{\ulTxSymbol}{\mbf{s}}
\newcommand{\ulTxData}{\mbf{x}}

\newcommand{\cellSharingPilotSetTp}[1]{\mathcal{L}_{#1}\lrc{\pilotReuseRatioTp}}
\newcommand{\cellSharingPilotSetSp}[1]{\mathcal{L}_{#1}\lrc{\pilotReuseRatioSp}}

\newcommand{\dlTxData}{d}
\newcommand{\dlSnr}{\gamma}
\newcommand{\dlNoise}{w}
\newcommand{\cpDlSinr}{\mathrm{SINR}^{\mathrm{RP-dl}}}
\newcommand{\cpDlRate}{R^{\mathrm{RP-dl}}}
\newcommand{\spDlRate}{R^{\mathrm{SP-dl}}}
\newcommand{\stagDlRate}{R^{\mathrm{ST-dl}}}

\newcommand{\spDlSinr}{\mathrm{SINR}^{\mathrm{SP-dl}}}

\newcommand{\cpMse}{\mathrm{NMSE}^{\mathrm{RP}}}
\newcommand{\spMse}{\mathrm{NMSE}^{\mathrm{SP}}}

\newcommand{\stagPDataPower}{p_d}
\newcommand{\stagPPilotPower}{p_p}
\newcommand{\stagDlSinr}{\mathrm{SINR}^{\mathrm{ST-dl}}}

\newcommand{\cpUlSinr}{\mathrm{SINR}^{\mathrm{RP-ul}}}

\newcommand{\hybridCpUlIci}{I^{\mathrm{RP-ul}}}
\newcommand{\hybridCpDlIci}{I^{\mathrm{RP-dl}}}
\newcommand{\hybridSpUlIci}{I^{\mathrm{SP-ul}}}
\newcommand{\hybridSpDlIci}{I^{\mathrm{SP-dl}}}
\newcommand{\hybridUlCostFactor}{\xi^{\mathrm{ul}}}
\newcommand{\hybridDlCostFactor}{\xi^{\mathrm{dl}}}
\newcommand{\hybridCpCost}{T^{\mathrm{RP}}}
\newcommand{\hybridSpCost}{T^{\mathrm{SP}}}
\newcommand{\hybridTotalIci}{I}
\newcommand{\hybridCompleteUserSet}{\mathcal{U}}

\newcommand{\cpPilot}{\pmb{\phi}}

\newcommand{\cardinality}[1]{\mathrm{card}\lrf{#1}}

\newcommand{\rhoD}[2]
{
	\ifstrempty{#1}
		{
			\ifstrempty{#2}
				{
					\rho
				}
				{
					\rho_{#2}
				}
		}
		{
			\ifstrempty{#2}
				{
					\rho_{#1}
				}
				{
					\rho_{#1,#2}
				}				
		}
}

\newcommand{\rhoP}[2]
{
	\ifstrempty{#1}
		{
			\ifstrempty{#2}
				{
					\lambda
				}
				{
					\lambda_{#2}
				}
		}
		{
			\ifstrempty{#2}
				{
				    \lambda_{#1}
				}
				{	
					\lambda_{#1,#2} 
				}
		}
}

\newcommand{\ulTotPower}[2]
{
	\ifstrempty{#1}
		{
			\ifstrempty{#2}
				{
					\mu
				}
				{
					\mu_{#2}
				}
		}
		{
			\mu_{#1#2}
		}
}

\newcommand{\spPilot}[2]
{
	\ifstrempty{#1}
		{
			\ifstrempty{#2}
				{
					\mbf{p}
				}
				{
					\mbf{p}_{#2}
				}
		}
		{
			\ifstrempty{#2}
				{
					\mbf{p}_{#1}
				}
				{
					\mbf{p}_{#1#2}
				}				
		}
}

\newcommand{\txPrecoder}{\mbf{g}}
\newcommand{\txPrecoderNormalization}{\nu}
\newcommand{\dlTxPower}{q}

\newcommand{\tTermOne}{t_1}
\newcommand{\tTermTwo}{t_2}

\newcommand{\userIdx}{n}
\newcommand{\symbolIdx}{t}

\newcommand{\pilotSharingSet}{\mathcal{P}}
\newcommand{\cValOne}{\alpha_1}
\newcommand{\cValTwo}{\alpha_2}
\newcommand{\cValThree}{\alpha_3}
\newcommand{\cValFour}{\alpha_4}
\newcommand{\cValFive}{\alpha_5}
\newcommand{\cValSix}{\alpha_6}
\newcommand{\cValSeven}{\alpha_7}

\newcommand{\signalPower}{s}
\newcommand{\interferencePower}{i}

\newcommand{\logCapacity}[1]{\Psi\lrc{#1}}
\newcommand{\userSet}{\Omega}

\newcommand{\tTerm}{v}

\newcommand{\rhoDrhoPConstant}{\kappa}

%% file: texFiles/abstract.tex
\begin{abstract}
	In this paper, we investigate the downlink throughput performance of a massive multiple-input multiple-output (MIMO) system that employs superimposed pilots for channel estimation. The component of downlink (DL) interference that results from transmitting data alongside pilots in the uplink (UL) is shown to decrease at a rate proportional to the square root of the number of antennas at the BS. The normalized mean-squared error (NMSE) of the channel estimate is compared with the Bayesian Cram\'{e}r-Rao lower bound that is derived for the system, and the former is also shown to diminish with increasing number of antennas at the base station (BS). Furthermore, we show that staggered pilots are a particular case of superimposed pilots and offer the downlink throughput of superimposed pilots while retaining the UL spectral and energy efficiency of regular pilots. We also extend the framework for designing a hybrid system, consisting of users that transmit either regular or superimposed pilots, to minimize both the UL and DL interference. The improved NMSE and DL rates of the channel estimator based on superimposed pilots are demonstrated by means of simulations. 
\end{abstract}


%% file: texFiles/introduction.tex
\section{Introduction}
Massive multiple-input multiple-output (MIMO) communication systems employ base stations (BS) with a large number of antennas and have garnered significant interest in recent years as a candidate for future fifth generation (5G) cellular systems \cite{boccardi2014fivedisruptive,jeffandrews2014whatwill,jungnickel2014theroleof,larsson2014massive,osseiran2014scenarios}. These systems promise a logarithmic increase in the uplink (UL) spectral and energy efficiency, with respect to the number of antennas at the BS, when the exact channel state information is assumed to be available for designing the detector \cite{ngo2013energy}. In practice, the channel state information (CSI) has to be obtained at the BS using orthogonal pilot sequences that are transmitted by the users. These orthogonal sequences typically occupy a portion of the time-frequency resource dedicated for pilot transmission (henceforth referred to as regular pilots (RP)). The overhead for obtaining the CSI increases linearly in the number of orthogonal pilot sequences transmitted, and therefore, in order to limit this overhead, pilot sequences are shared/reused across cells in multi-cell systems. This sharing results in inter-cell interference in both the UL and downlink (DL), which is known as \textit{pilot contamination} \cite{Marzetta2010Noncooperative}. Pilot contamination diminishes the promised gains of massive MIMO systems and hence is considered a major impediment \cite{lulu2014anoverview}. Approaches for pilot decontamination have garnered significant interest in recent years  and they primarily rely on separating the users based on properties such as asymptotic orthogonality between user channels, non-overlapping angle of arrivals of the signal at the BS,  and pilot reuse  \cite{Muller2014Blind,ngo2012evd,bjornson2015massive,upadhya2015anarray,Yin2013Coordinated,bjornson2017pilot}.

Superimposed pilots (SP) have been extensively studied for channel estimation in MIMO systems \cite{takeuchi2013Achievable,shuangchi2007superimposed,coldrey2007training,haidong2003pilot,cui2005pilot}, especially in the context of rapidly changing channels in which reserving a set of symbols for pilot transmission would be impractical. Recently,
in \cite{upadhya2016superimposed}, SPs have been studied as an alternative pilot structure to mitigate/avoid pilot contamination in massive MIMO. SPs have also been investigated for use in massive MIMO systems in \cite{verenzuela2017spectral,Zhang2015SuperimposedPilot}. In \cite{verenzuela2017spectral}, the authors derive expressions for the UL spectral and energy efficiency of SP and compare them with those for RP. In \cite{Zhang2015SuperimposedPilot}, the authors have considered the case when the number of symbols in the UL time-slot is larger than the number of users in the system. 

In \cite{upadhya2016superimposed}, approximate expressions have been derived for the UL signal-to-interference-plus-noise ratio (SINR) and rate at the output of a matched filter (MF) that employs the least-squares (LS)-based channel estimate in an iterative and non-iterative way. The expressions have been derived under the condition that the total number of users in the system is smaller than the number of symbols in the UL time-slot. The importance of power control for a system employing SP has been highlighted and the fractions of power that should be assigned to pilots and data, respectively, in order to maximize an approximation on the UL per cell rate, have been derived. It has been found that with increasing number of antennas at the BS, the optimal fraction of the power assigned to the data would decrease proportional to the square root of the number of antennas at the BS. In addition, a hybrid system that employs both RP and SP has been introduced to minimize the total UL interference, and shown to be superior to a system that is optimized for maximal spectral efficiency \cite{bjornson2015massive} but employs only RP. 

In this paper, we provide additional important theoretical results with regard to SP for massive MIMO systems through performance metrics such as the normalized mean-squared error (NMSE) of the channel estimate and especially the DL rate. In particular, the following are the contributions of this paper.
\begin{itemize}
	\item Closed-form expressions for the DL achievable rate are derived when the channel estimates obtained from SP are employed in a MF precoder at the BS. 
	\item We discuss the relationship between staggered pilots and SP and derive the DL rate for the former scheme.
	\item We derive expressions for the NMSE and compare it against the Bayesian Cram\'er-Rao lower bound (CRLB) that we also derive for the system.
	\item The hybrid system described in \cite{upadhya2016superimposed}, which consists of users that transmit both RP and SP, is extended to the DL and is designed by minimizing both the UL and DL interference.
	\item Simulations are carried out to validate the MSE and DL performance of SP and the hybrid system.
\end{itemize}
Some initial results, for the CRLB and approximate DL rate, have been reported in \cite{upadhya2016downlink} without detailed derivations. In addition, some results for the hybrid system, with approximate UL and DL rates, have been reported in \cite{upadhya2016hybrid}.

The paper is organized as follows. In Section \ref{sec:systemModel}, we briefly review the system model for the UL and introduce the system model for the DL. In Section \ref{sec:pilotContaminationDownlink}, the DL rate is derived when the channel estimates are employed in an MF precoder. In addition, the expressions for the MSE and the corresponding CRLB of the channel estimate are derived for a system that employs SP. These metrics are then compared with the corresponding metrics that are obtained for a system employing RP. In Section \ref{sec:staggeredPilots}, staggered pilots are shown to be a particular case of SP and the DL rate for this scheme is derived. In Section \ref{sec:hybridSystem}, the framework for the hybrid system proposed in \cite{upadhya2016superimposed} is extended to include the downlink. Using simulations, Section \ref{sec:simulationResults} discusses the performance of the hybrid system and compares the MSE and DL performance of RP and SP. Section \ref{sec:conclusion} concludes the paper. Some of the lengthy proofs and derivations are detailed in the appendix.

\textit{Notation} : Lower case and upper case boldface letters denote column vectors and matrices, respectively. The notations $(\cdot)^*,(\cdot)^T$, $ (\cdot)^H $, and $ (\cdot)^{-1} $ represent the conjugate, transpose, Hermitian transpose, and inverse, respectively. The Kronecker product is denoted by $ \otimes $. The notation $ \mathcal{C}\mathcal{N}(\pmb{\mu},\mbf{\Sigma}) $ stands for the complex normal distribution with mean $ \pmb{\mu} $ and covariance matrix $ \mbf{\Sigma} $ and $ \mathbb{E}\lrf{\cdot} $ denotes the expectation operator. The notation $ \mathbf{I}_{N} $ is used to denote an $ N\times N $ identity matrix and $ \|\cdot\| $ denotes the Euclidean norm of a vector. Upper case calligraphic letters denote sets and $ \indicator{\mathcal{S}} $ represents the indicator function over the set $ \mathcal{S} $, while $ \delta_{n,m} $ denotes the Kronecker delta function. The empty set is denoted by $ \varnothing $, whereas the symbols $ \cup $~and~$ \backslash $ stands for the union and the relative complement operations, respectively. The operator $ \lfloor x \rfloor $ returns the largest integer smaller than $ x $. The trace of matrix $ \mbf{A} $ is represented as $ \mathrm{trace}\lrf{\mbf{A}} $.

%% file: texFiles/systemModel.tex
\section{System Model}
\label{sec:systemModel}

We consider a time-division duplexing (TDD) massive MIMO system with $ L $ cells and $ K $ single-antenna users per cell. Each cell has a BS with $ M $ antennas. In the UL phase, the users transmit $ \ulDuration $ symbols, which include both data and pilots. Using the tuple $ \lrc{\sIdxOne,\sIdxTwo} $ to denote user $ \sIdxTwo $ in cell~$ \sIdxOne $, the matrix of received symbols $ \mbf{Y}_{\vIdxOne}\in\mathbb{C}^{M\times\ulDuration} $ at BS $ \vIdxOne $ can be written as
\begin{equation}
	\mbf{Y}_{\vIdxOne} = \sum\limits_{\sIdxOne=0}^{L-1} \sum\limits_{\sIdxTwo=0}^{K-1} \sqrt{\ulTotPower{\sIdxOne}{\sIdxTwo}} \mbf{h}_{\vIdxOne\sIdxOne\sIdxTwo} \ulTxSymbol_{\sIdxOne\sIdxTwo}^T + \mbf{W}_{\vIdxOne}
	\label{eqn:ulSystemModel}
\end{equation}
where $ \mbf{h}_{\vIdxOne\sIdxOne\sIdxTwo} \in \mathbb{C}^{M} $ is the channel response between BS $ \vIdxOne $ and user $ \lrc{\sIdxOne,\sIdxTwo} $, $ \ulTxSymbol_{\sIdxOne\sIdxTwo} \in \mathbb{C}^{\ulDuration} $ is the vector of symbols transmitted by user $ \lrc{\sIdxOne,\sIdxTwo} $ with power $ \ulTotPower{\sIdxOne}{\sIdxTwo} $, and $ \mbf{W}_{\vIdxOne} \in \mathbb{C}^{M\times\ulDuration} $ is the matrix of additive white Gaussian noise at BS $ \vIdxOne $ with each column distributed as $ \mathcal{CN}(\mathbf{0},\sigma^2\eye) $ and being mutually independent of the other columns. The channel vectors $ \mbf{h}_{\vIdxOne\sIdxOne\sIdxTwo} $ are assumed to be distributed as $ \mathcal{CN}(\mathbf{0},\beta_{\vIdxOne\sIdxOne\sIdxTwo}\eye) $ where $ \beta_{\vIdxOne\sIdxOne\sIdxTwo} $ denotes the large-scale path-loss coefficient. In addition, the channel is assumed to be constant during the coherence time, i.e., $ \coherenceTime $ symbols whereas $ \beta_{\vIdxOne\sIdxOne\sIdxTwo} $ is constant for a significantly longer duration than $ \coherenceTime $ symbols. 

The symbol $ \ulTxSymbol_{\sIdxOne\sIdxTwo} $ is dependent on the nature of the pilot transmitted in the UL. For example, when RP is employed, a part of $ \ulTxSymbol_{\sIdxOne\sIdxTwo} $ is reserved for pilots and the remaining part is used for data transmission. Whereas, when SP is employed, the whole of $ \ulTxSymbol_{\sIdxOne\sIdxTwo} $ contains both pilots and data.

Assuming channel reciprocity, if the BS $ \sIdxOne $ uses the precoder $ \txPrecoder_{\sIdxOne\sIdxTwo} $ and if $ \dlTxData_{\sIdxOne\sIdxTwo} \in\mathbb{C} $ is the data symbol transmitted to user $ \lrc{\sIdxOne,\sIdxTwo} $ by BS $ \sIdxOne $, then the received symbol at user $ \lrc{\vIdxOne,\vIdxTwo} $ can be written as
\begin{equation}
	\widehat{\dlTxData}_{\vIdxOne\vIdxTwo} 
	= 	
	\sqrt{\dlSnr}
	\sum\limits_{\sIdxOne=0}^{L-1} 
	\mbf{h}_{\sIdxOne\vIdxOne\vIdxTwo}^H 
	\sum\limits_{\sIdxTwo=0}^{K-1}
	\sqrt{\txPrecoderNormalization_{\sIdxOne\sIdxTwo}}
	\txPrecoder_{\sIdxOne\sIdxTwo}	 
	\dlTxData_{\sIdxOne\sIdxTwo} 
	+ 
	\dlNoise_{\vIdxOne\vIdxTwo}
	\label{eqn:systemModelDownlink}
\end{equation}
where $ \dlSnr $ is the DL SNR of user $ \lrc{\vIdxOne,\vIdxTwo} $ and is assumed to be same in all the cells, and $ \dlNoise_{\vIdxOne,\vIdxTwo} $ is zero-mean unit-variance additive Gaussian noise at the user terminal. The symbols $ \dlTxData_{\sIdxOne\sIdxTwo}, \forall \; \lrc{\sIdxOne,\sIdxTwo} $ are assumed to be distributed as $ \dlTxData_{\sIdxOne\sIdxTwo} \sim \mathcal{CN}\lrc{0,1} $ and are statistically independent of the channel vectors $ \mbf{h} $ and the UL symbols $ \ulTxSymbol $. The parameter $ \txPrecoderNormalization_{\sIdxOne\sIdxTwo} = \dlTxPower_{\sIdxOne\sIdxTwo} / \expectation\lrf{\|\txPrecoder_{\sIdxOne\sIdxTwo}\|^2} $ normalizes the average transmit power to user $ \lrc{\sIdxOne,\sIdxTwo} $ to be $ \dlTxPower_{\sIdxOne\sIdxTwo} $ \cite{hoydis2013massive,bjornson2015massive}.

%% file: texFiles/pilotContaminationDownlink.tex
\section{Effect of Pilot Contamination on the Downlink}
\label{sec:pilotContaminationDownlink}
In TDD massive MIMO, under the assumption of channel reciprocity, the precoder for data transmission in the DL is designed using the channel estimate that is obtained from UL training. Therefore, the throughput in the DL depends on the quality of the channel that has been estimated in the UL. In this section, the quality of the channel estimate obtained from both RP and SP transmission schemes are quantified through the normalized MSE and the latter is compared with the CRLB. In addition, the closed form expressions for the  DL achievable rate at the user terminal are derived and compared when the channel estimates are used in an MF precoder. 
\subsection{Regular Pilots}
\label{sec:pilotContaminationDownlinkTp}
With RP, each user transmits a $ \tau \geq K $ length pilot sequence for channel estimation followed by UL data. Let the length-$ \tau $ pilot sequences be taken from the columns of a scaled unitary matrix $ \mbf{\Phi} \in \mathbb{C}^{\tau\times\tau} $ such that $ \mbf{\Phi}^H\mbf{\Phi} = \tau\eye_{\tau} $. These orthogonal pilot sequences are distributed across $ \pilotReuseRatioTp \triangleq \lfloor\tau/K\rfloor $ cells, where $ \pilotReuseRatioTp $ is assumed to be a positive integer. In other words, the pilot sequence $ \pmb{\phi}_{\sIdxOne\sIdxTwo} $ that is transmitted by user $ \lrc{\sIdxOne,\sIdxTwo}$ is reused at every $ \pilotReuseRatioTp\rth $ cell.
Assuming that all the pilot transmissions are synchronized, the LS estimate of the channel can be easily found as \cite{Marzetta2010Noncooperative,upadhya2016superimposed}
\begin{equation}
	\label{eqn:lsConventionalPilotChannelEstimate}
	\widehat{\mbf{h}}_{\vIdxOne\vIdxOne\vIdxTwo}^{\mathrm{RP}} 
	= 
	\mbf{h}_{\vIdxOne\vIdxOne\vIdxTwo} 
	+  
	\sum\limits_{\mathcal{L}_{\vIdxOne}(\pilotReuseRatioTp)\ni\sIdxOne\neq\vIdxOne} 
	\sqrt{\frac{\ulTotPower{\sIdxOne}{\vIdxTwo}}{\ulTotPower{\vIdxOne}{\vIdxTwo}}}
	\mbf{h}_{\vIdxOne\sIdxOne\vIdxTwo} + 
	\mbf{w}_{\vIdxOne\vIdxTwo} 
\end{equation}
where $ \mbf{w}_{\vIdxOne\vIdxTwo} = \mbf{W}_{\vIdxOne}\pmb{\phi}_{\vIdxOne\vIdxTwo}^*/\lrc{\tau\sqrt{\ulTotPower{\vIdxOne}{\vIdxTwo}}} $ and  $ \mathcal{L}_{\vIdxOne}(\pilotReuseRatioTp) $ is the subset of the $ L $ cells that use the same pilot sequences as cell $ \vIdxOne $. The normalized MSE of the channel estimate $ \widehat{\mbf{h}}_{\vIdxOne\vIdxOne\vIdxTwo}^{\mathrm{RP}}  $ is defined as
\begin{align}
		\cpMse_{\vIdxOne\vIdxTwo} 
		\!
		&\triangleq
		\frac{
		\expectation
		\lrf{
			\|
			\widehat{\mbf{h}}_{\vIdxOne\vIdxOne\vIdxTwo}^{\mathrm{RP}}
			-
			\mbf{h}_{\vIdxOne\vIdxOne\vIdxTwo}
			\|^2
		}
		}{\expectation\lrf{\|\mbf{h}_{\vIdxOne\vIdxOne\vIdxTwo}\|^2}}		
	= 
	\frac{1}{\beta_{\vIdxOne\vIdxOne\vIdxTwo}}
	\lrc{
	\sum\limits_{\mathcal{L}_{\vIdxOne}(\pilotReuseRatioTp)\ni\sIdxOne\neq\vIdxOne} 
	\frac{\ulTotPower{\sIdxOne}{\vIdxTwo}}{\ulTotPower{\vIdxOne}{\vIdxTwo}}
	\beta_{\vIdxOne\sIdxOne\vIdxTwo} 
	+ 
	\frac{\sigma^2}{\tau\ulTotPower{\vIdxOne}{\vIdxTwo}}} \;.
	\label{eqn:cpMse}
\end{align}
The first term in \eqref{eqn:cpMse} is the estimation error due to pilot contamination from users in the neighboring cells which employ the same pilots as user $ \lrc{\vIdxOne,\vIdxTwo} $. 

If $ \dlDuration $ symbols are transmitted from the BS to the user terminals in the DL phase, then the rate in the downlink for user $ \lrc{\vIdxOne,\vIdxTwo} $ can be expressed as \cite{Marzetta2010Noncooperative}
\begin{equation}
\cpDlRate_{\vIdxOne\vIdxTwo} 
= 
\frac{\dlDuration}{\coherenceTime}\log_2\lrc{1 + \cpDlSinr_{\vIdxOne\vIdxTwo}}
\label{eqn:cpDlRate}
\end{equation}
where $ \coherenceTime = \ulDuration + \dlDuration $ is the smallest channel coherence time of all the users in the system, and $ \cpDlSinr_{\vIdxOne\vIdxTwo} $ is the DL SINR at user $ \lrc{\vIdxOne,\vIdxTwo} $. If the channel estimate in \eqref{eqn:lsConventionalPilotChannelEstimate} is used in an MF precoder, then \cite{jose2011pilot}
\begin{align}
\cpDlSinr_{\vIdxOne\vIdxTwo}
=
\frac{
	\txPrecoderNormalization_{\vIdxOne\vIdxTwo}
	\beta_{\vIdxOne\vIdxOne\vIdxTwo}^2
}
{
	\!\!\!
	\sum\limits_{\cellSharingPilotSetTp{\vIdxOne}\ni\sIdxOne\neq\vIdxOne}
	\frac{\ulTotPower{\vIdxOne}{\vIdxTwo}}{\ulTotPower{\sIdxOne}{\vIdxTwo}}
	\txPrecoderNormalization_{\sIdxOne\vIdxTwo}
	\beta_{\sIdxOne\vIdxOne\vIdxTwo}^2
	+
	\!
	\frac{1}{M}
	\!
	\sum\limits_{\sIdxOne=0}^{L-1}
	\sum\limits_{\sIdxTwo=0}^{K-1}
	\frac{\txPrecoderNormalization_{\sIdxOne\sIdxTwo}}{\ulTotPower{\sIdxOne}{\sIdxTwo}}
	\!
	\lrc{
		\!
	\sum\limits_{\sIdxThree\in\cellSharingPilotSetTp{\sIdxOne}}	
	\!\!
	\ulTotPower{\sIdxThree}{\sIdxTwo}
	\beta_{\sIdxOne\vIdxOne\vIdxTwo}
	\beta_{\sIdxOne\sIdxThree\sIdxTwo}
	\!
	+
	\!
	\frac{\sigma^2\beta_{\sIdxOne\vIdxOne\vIdxTwo}}{\tau}
	\!\!
	}
	+
	\frac{1}{M^2\dlSnr}
}
\label{eqn:cpDlSinrMf}
\end{align}
where
\begin{align}
	\txPrecoderNormalization_{\sIdxOne\sIdxTwo}
	=
	\frac{\dlTxPower_{\sIdxOne\sIdxTwo}}{\expectation\lrf{\|\txPrecoder_{\sIdxOne\sIdxTwo}\|^2}}
	=
	\frac{\dlTxPower_{\sIdxOne\sIdxTwo}}{M}
	\lrc{
		\sum\limits_{\sIdxThree\in\cellSharingPilotSetTp{\sIdxOne}}
		\frac{\ulTotPower{\sIdxThree}{\sIdxTwo}}{\ulTotPower{\sIdxOne}{\sIdxTwo}}
		\beta_{\sIdxOne\sIdxThree\sIdxTwo}
		+
		\frac{\sigma^2}{\tau\ulTotPower{\sIdxOne}{\sIdxTwo}}
		}^{-1}\;.
\end{align}
The estimation error due to pilot contamination limits the asymptotic $ \lrc{M\rightarrow\infty} $ DL SINR of user $ \lrc{\vIdxOne,\vIdxTwo} $ to \cite{Marzetta2010Noncooperative}
\begin{equation}
	\cpDlSinr_{\vIdxOne\vIdxTwo} = \frac{\widetilde{\txPrecoderNormalization}_{\vIdxOne\vIdxTwo}\beta_{\vIdxOne\vIdxOne\vIdxTwo}^2}{\sum\limits_{\mathcal{L}_{\vIdxOne}(\pilotReuseRatioTp)\ni\sIdxOne\neq\vIdxOne}\widetilde{\txPrecoderNormalization}_{\sIdxOne\vIdxTwo}\beta_{\sIdxOne\vIdxOne\vIdxTwo}^2} \ .
	\label{eqn:cpDlSinrMfInfM}
\end{equation}
where $ \widetilde{\txPrecoderNormalization}_{\sIdxOne\sIdxTwo} = M \txPrecoderNormalization_{\sIdxOne\sIdxTwo} $. Note that $ \widetilde{\txPrecoderNormalization}_{\sIdxOne\sIdxTwo} $ is independent of $ M $.

\subsection{Superimposed Pilots}
When employing SP, the estimate of the channel is obtained from pilots that are transmitted at a reduced power alongside the data. The LS estimate of the channel can be written as \cite{upadhya2016superimposed} 
\begin{align}
\label{eqn:superimposedPilotNonIterativeChannel}
\widehat{\mbf{h}}_{\vIdxOne\sIdxOne\sIdxTwo}^{\mathrm{SP}}
&= 
\sum_{\sIdxThree\in\cellSharingPilotSetSp{\vIdxOne}}
\sqrt{\frac{\ulTotPower{\sIdxThree}{\sIdxTwo}}{\ulTotPower{\sIdxOne}{\sIdxTwo}}}
\mbf{h}_{\vIdxOne\sIdxThree\sIdxTwo}
+
\frac{\rhoD{}{}}{\ulDuration\rhoP{}{}} 
\sum\limits_{\sIdxThree=0}^{L-1}
\sum\limits_{\sIdxFour=0}^{K-1}
\sqrt{\frac{\ulTotPower{\sIdxThree}{\sIdxFour}}{\ulTotPower{\sIdxOne}{\sIdxTwo}}}
\mbf{h}_{\vIdxOne\sIdxThree\sIdxFour} 
\ulTxData_{\sIdxThree\sIdxFour}^T
\spPilot{\sIdxOne}{\sIdxTwo}^*  
+
\frac{\mbf{W}_{\vIdxOne}\spPilot{\sIdxOne}{\sIdxTwo}^*}{\ulDuration\rhoP{}{}\sqrt{\ulTotPower{\sIdxOne}{\sIdxTwo}}} 
\end{align}
where $ \spPilot{\vIdxOne}{\vIdxTwo} \in \mathbb{C}^{\ulDuration} $ and $ \ulTxData_{\vIdxOne\vIdxTwo} \in \mathbb{C}^{\ulDuration}  $ are, respectively, the pilot and data vectors transmitted by user $ \lrc{\vIdxOne,\vIdxTwo} $, $ \pilotReuseRatioSp \triangleq \lfloor\ulDuration/K\rfloor $ is a positive integer representing the number of cells over which SP are reused, $ \cellSharingPilotSetSp{\vIdxOne} $ is the subset of the $ L $ cells that use the same pilot sequences as cell $ \vIdxOne $. In addition, the pilots are taken from the columns of a scaled unitary matrix $ \mbf{P} \in \mathbb{C}^{\ulDuration\times\ulDuration} $ such that $ \mbf{P}^H\mbf{P} = \ulDuration\eye_{\ulDuration} $, and therefore, $ \spPilot{\sIdxOne}{\sIdxTwo}^H\spPilot{\sIdxThree}{\sIdxFour} = \ulDuration \delta_{\sIdxOne\sIdxThree} \delta_{\sIdxTwo\sIdxFour} $. The parameters $ \rhoP{}{}^2> 0 $  and $ \rhoD{}{}^2> 0 $ are the fractions of the UL transmit power reserved for pilots and data, respectively, such that $ \rhoP{}{}^2 + \rhoD{}{}^2 = 1 $. Moreover, in \eqref{eqn:superimposedPilotNonIterativeChannel}, it is assumed that every user in the system uses the same value of $ \rhoP{}{} $ and $ \rhoD{}{} $. 

Similar to \eqref{eqn:cpMse}, the normalized MSE for the channel estimate obtained from SP is defined as
\begin{align}
	\spMse_{\vIdxOne\vIdxTwo} 
	\!
	&\triangleq 
	\!
	\frac{
	\expectation
	\lrf{		
		\|
		\widehat{\mbf{h}}^{\mathrm{SP}}_{\vIdxOne\vIdxOne\vIdxTwo}
		-
		\mbf{h}_{\vIdxOne\vIdxOne\vIdxTwo}
		\|^2	
		}}
		{\expectation\lrf{\|\mbf{h}_{\vIdxOne\vIdxOne\vIdxTwo}\|^2}}
\!
\nonumber
\\
&=
\!
\frac{1}{\beta_{\vIdxOne\vIdxOne\vIdxTwo}}
\lrc{
\sum\limits_{\cellSharingPilotSetSp{\vIdxOne}\ni\sIdxOne\neq\vIdxOne}
\frac{\ulTotPower{\sIdxOne}{\vIdxTwo}}{\ulTotPower{\vIdxOne}{\vIdxTwo}}
\beta_{\vIdxOne\sIdxOne\vIdxTwo}
+
\frac{\rhoD{}{}^2}{\ulDuration\rhoP{}{}^2} 
\sum\limits_{\sIdxOne=0}^{L-1}
\sum\limits_{\sIdxTwo=0}^{K-1}
\frac{\ulTotPower{\sIdxOne}{\sIdxTwo}}{\ulTotPower{\vIdxOne}{\vIdxTwo}}
\beta_{\vIdxOne\sIdxOne\sIdxTwo}
+
\frac{\sigma^2}{\rhoP{}{}^2\ulDuration\ulTotPower{\vIdxOne}{\vIdxTwo}}} \;.
\label{eqn:spMse}
\end{align}
The first error term in \eqref{eqn:spMse} results from reusing pilots every $ \pilotReuseRatioSp $ cells, whereas the second error term results from transmitting pilots alongside data. As in the case of RP, both the errors lead to interference in the DL phase. Under the assumption that the interference from outside the $ \pilotReuseRatioSp $ contiguous cells which contain the reference BS can be neglected, the CRLB for the channel estimate can be derived as (the derivation is in Appendix \ref{appdx:MSE})
\begin{align}
\mathrm{CRLB}\lrc{\mbf{h}_{\vIdxOne\vIdxOne\vIdxTwo}} 
=
\frac{M}{\frac{\ulDuration}{\sigma^2}+\frac{1}{\ulTotPower{\vIdxOne}{\vIdxTwo}\beta_{\vIdxOne\vIdxOne\vIdxTwo}}}
\approx
\frac{M\sigma^2}{\ulDuration}
\label{eqn:spCrlb}
\end{align}
where the approximation is valid when $ \sigma^2/\ulDuration \ll \ulTotPower{\vIdxOne}{\vIdxTwo}\beta_{\vIdxOne\vIdxOne\vIdxTwo} $. Therefore, we have the relation,
\begin{align}
	\spMse_{\vIdxOne\vIdxTwo} \geq \frac{1}{M\beta_{\vIdxOne\vIdxOne\vIdxTwo}} \mathrm{CRLB}\lrc{\mbf{h}_{\vIdxOne\vIdxOne\vIdxTwo}} 
	\approx \frac{\sigma^2}{\beta_{\vIdxOne\vIdxOne\vIdxTwo}\ulDuration}\;.
\end{align}
The Bayesian CRLB is a lower bound on the MSE of a minimum mean-squared error (MMSE) channel estimator, and its value in \eqref{eqn:spCrlb} is the MSE of an MMSE estimator when $ \rhoD{}{} = 0 $ or all the power is allocated to the pilots. In addition, the approximation in \eqref{eqn:spCrlb} is the MSE of the LS estimator when $ \rhoD{}{} = 0 $.

The NMSE in \eqref{eqn:spMse} is parameterized by both $ \rhoD{}{}^2 $ and $ \rhoP{}{}^2 $. However, the CRLB is loose for non-zero values of $ \rhoD{}{}^2 $ and an estimator will attain this bound only when $ \rhoD{}{}^2 = 0 $, i.e., when all the power is allocated to the pilots. Nevertheless, the CRLB is a standard, as well as, a useful benchmark to evaluate the performance of the proposed method. For example, in the same context of massive MIMO, the performance of a semi-blind channel estimation method is also compared against the CRLB in \cite{nayebi2018semi-blind}. We will also see in Section \ref{sec:simulationResults} that the CRLB in \eqref{eqn:spCrlb} is achieved\footnote{In fact, it is the approximation in \eqref{eqn:spCrlb} that is achieved by the estimator.} by the estimator in \eqref{eqn:superimposedPilotNonIterativeChannel} when $ M \rightarrow \infty $.

A lower bound on the DL ergodic capacity can be obtained for superimposed pilots using a similar approach as in \cite{jose2011pilot}. Rewriting \eqref{eqn:systemModelDownlink} as
\begin{align}
\widehat{\dlTxData}_{\vIdxOne\vIdxTwo}
&= 
\sqrt{\dlSnr
	\txPrecoderNormalization_{\vIdxOne\vIdxTwo}}
\expectation
\lrf{
	\mbf{h}_{\vIdxOne\vIdxOne\vIdxTwo}^H
	\txPrecoder_{\vIdxOne\vIdxTwo}
}
\dlTxData_{\vIdxOne\vIdxTwo} 
+
\sqrt{\dlSnr
	\txPrecoderNormalization_{\vIdxOne\vIdxTwo}}
\lrc{	
	\mbf{h}_{\vIdxOne\vIdxOne\vIdxTwo}^H 
	\txPrecoder_{\vIdxOne\vIdxTwo} 
	- 
	\expectation
	\lrf{
		\mbf{h}_{\vIdxOne\vIdxOne\vIdxTwo}^H
		\txPrecoder_{\vIdxOne\vIdxTwo}
	}	
}
\dlTxData_{\vIdxOne\vIdxTwo}
\nonumber
\\
&\qquad+
\sqrt{\dlSnr}
\mathop
{
	\sum
	\sum
}_{\lrc{\sIdxOne,\sIdxTwo}\neq\lrc{\vIdxOne,\vIdxTwo}}
\sqrt{\txPrecoderNormalization_{\sIdxOne\sIdxTwo}}
\mbf{h}_{\sIdxOne\vIdxOne\vIdxTwo}^H
\txPrecoder_{\sIdxOne\sIdxTwo}	
\dlTxData_{\sIdxOne\sIdxTwo}
+
w_{\vIdxOne\vIdxTwo} \;,
\label{eqn:dlDataTxn}
\end{align}
and noting that the first term is uncorrelated with the subsequent terms, a lower bound on the ergodic capacity can be computed as \cite{Hassibi2003HowMuch}
\begin{align}
\spDlRate_{\vIdxOne\vIdxTwo} 
=
\frac{\dlDuration}{\coherenceTime}
\expectation
\lrf{
\log_2
\lrc{1 + \spDlSinr_{\vIdxOne\vIdxTwo}}
}
\label{eqn:spDlRate}
\end{align}
where
\begin{align}
\spDlSinr_{\vIdxOne\vIdxTwo}
=
\frac{
	\txPrecoderNormalization_{\vIdxOne\vIdxTwo}
	|
	\expectation
	\lrf{			
		\mbf{h}_{\vIdxOne\vIdxOne\vIdxTwo}^H
		\txPrecoder_{\vIdxOne\vIdxTwo}
	}
	|^2
	}
{
\sum_{\sIdxOne=0}^{L-1}
\sum_{\sIdxTwo=0}^{K-1}
\txPrecoderNormalization_{\sIdxOne\sIdxTwo}
\expectation
\lrf{
	|
	\mbf{h}_{\sIdxOne\vIdxOne\vIdxTwo}^H
	\txPrecoder_{\sIdxOne\sIdxTwo}	
	|^2}
-
\txPrecoderNormalization_{\vIdxOne\vIdxTwo}
|
\expectation
\lrf{			
	\mbf{h}_{\vIdxOne\vIdxOne\vIdxTwo}^H
	\txPrecoder_{\vIdxOne\vIdxTwo}
}
|^2
+
\frac{1}{\dlSnr}
}\;.
\label{eqn:interferencePowerExpansion}
\end{align}
In \eqref{eqn:spDlRate} and \eqref{eqn:interferencePowerExpansion}, the expectation outside the logarithm is with respect to the user locations whereas the inner expectation is with respect to the channel and noise vectors. In addition, these expressions are valid for any combining scheme. However, to obtain closed-form expressions for precoders such as zero-forcing (ZF) or MMSE, we require the channel estimation error to be independent of the estimate. But, as outlined in \cite{verenzuela2017spectral}, the LS channel estimate and the estimation error when SP is employed are not Gaussian. As a result, even if a linear minimum mean-squared error (LMMSE) channel estimate were to be employed, it would only result in the estimation error being uncorrelated with the estimate but not independent of it. This renders it difficult/impossible in general to obtain closed-form expressions for precoders such as ZF and MMSE. We will therefore obtain a closed form expression for the SINR for MF precoding and numerically evaluate \eqref{eqn:spDlRate} and \eqref{eqn:interferencePowerExpansion} in Section \ref{sec:simulationResults} for methods such as ZF. Setting $ \txPrecoder_{\sIdxOne\sIdxTwo} =  \widehat{\mbf{h}}_{\sIdxOne\sIdxOne\sIdxTwo} $, the DL SINR for the MF precoder has been obtained in Appendix \ref{appdx:dlSinr} as
\begin{align}
\spDlSinr_{\vIdxOne\vIdxTwo}
\!
&=
\!
\txPrecoderNormalization_{\vIdxOne\vIdxTwo}\beta_{\vIdxOne\vIdxOne\vIdxTwo}^2
\!\!
\left(
\sum\limits_{\cellSharingPilotSetSp{\vIdxOne}\ni\sIdxOne\neq\vIdxOne}
\!
\frac{\ulTotPower{\vIdxOne}{\vIdxTwo}}{\ulTotPower{\sIdxOne}{\vIdxTwo}}
\txPrecoderNormalization_{\sIdxOne\vIdxTwo}
\beta_{\sIdxOne\vIdxOne\vIdxTwo}^2
+
\frac{1}{M}
\sum\limits_{\sIdxOne=0}^{L-1}
\sum\limits_{\sIdxTwo=0}^{K-1}
\frac{\txPrecoderNormalization_{\sIdxOne\sIdxTwo}
	\beta_{\sIdxOne\vIdxOne\vIdxTwo}}{\ulTotPower{\sIdxOne}{\sIdxTwo}}
\!\!
\lrc{
	\sum\limits_{\sIdxThree\in\cellSharingPilotSetSp{\sIdxOne}}		
	\ulTotPower{\sIdxThree}{\sIdxTwo}
	\beta_{\sIdxOne\sIdxThree\sIdxTwo}
	+
	\frac{\sigma^2}{\ulDuration\rhoP{}{}^2}
	\!
}
\right.
\nonumber
\\
&\left.
+
\frac{\rhoD{}{}^2}{\ulDuration\rhoP{}{}^2}
\sum\limits_{\sIdxOne=0}^{L-1}
\sum\limits_{\sIdxTwo=0}^{K-1}		
\frac{\txPrecoderNormalization_{\sIdxOne\sIdxTwo}}{\ulTotPower{\sIdxOne}{\sIdxTwo}}
\lrc{
	\ulTotPower{\vIdxOne}{\vIdxTwo}
	\beta_{\sIdxOne\vIdxOne\vIdxTwo}^2
	+
	\frac{1}{M}
	\sum\limits_{\sIdxThree=0}^{L-1}		
	\sum\limits_{\sIdxFour=0}^{K-1}
	\ulTotPower{\sIdxThree}{\sIdxFour}				
	\beta_{\sIdxOne\sIdxThree\sIdxFour}
	\beta_{\sIdxOne\vIdxOne\vIdxTwo}
}
+
\frac{1}{M^2\dlSnr}
\right)^{-1}
\label{eqn:spDlSinrMf}
\end{align}
where
\begin{equation}
	\txPrecoderNormalization_{\sIdxOne\sIdxTwo}
	=
	\frac{\dlTxPower_{\sIdxOne\sIdxTwo}}{\expectation
		\lrf{\|\txPrecoder_{\sIdxOne\sIdxTwo}\|^2}}
	=
	\frac{\dlTxPower_{\sIdxOne\sIdxTwo}\ulTotPower{\sIdxOne}{\sIdxTwo}}{M}
	\lrc{
		\sum\limits_{\sIdxThree\in\cellSharingPilotSetSp{\sIdxOne}}
		\ulTotPower{\sIdxThree}{\sIdxTwo}
		\beta_{\sIdxOne\sIdxThree\sIdxTwo}
		+
		\sum\limits_{\sIdxThree=0}^{L-1}	
		\sum\limits_{\sIdxFour=0}^{K-1}	
		\ulTotPower{\sIdxThree}{\sIdxFour}
		\frac{\rhoD{}{}^2}{\ulDuration\rhoP{}{}^2}
		\beta_{\sIdxOne\sIdxThree\sIdxFour}
		+
		\frac{\sigma^2}{\ulDuration\rhoP{}{}^2}
	}^{-1}\;.
\end{equation}

In \cite{upadhya2016superimposed}, the optimal values of $ \rhoD{}{}^2 $ and $ \rhoP{}{}^2 $ were computed by maximizing an approximation of the UL sum-rate which was obtained assuming that $ \ulDuration \geq LK $. 
However, since in practice $ \ulDuration \leq LK $, the results obtained in \cite{upadhya2016superimposed} do not necessarily hold. In this paper, we obtain the optimal values of $ \rhoD{}{}^2 $ and $ \rhoP{}{}^2 $ by maximizing an achievable lower bound on the channel capacity when $ \ulDuration \leq LK $.
\begin{prop}
	The values of $ \rhoD{}{}^2 $ and $ \rhoP{}{}^2 $ that maximize the achievable rate in the UL are 
	\begin{align}
		\rhoD{}{}^2_{\mathrm{opt}}
		&=
		\lrc{1 + \sqrt{M} \rhoDrhoPConstant}^{-1}
		\label{eqn:rhoDUlOptimal}
		\\
		\rhoP{}{}^2_{\mathrm{opt}}
		&=
		\lrc{1 + \frac{1}{\rhoDrhoPConstant\sqrt{M}}}^{-1}
		\label{eqn:rhoPUlOptimal}
	\end{align}
	where 
	\begin{align}
	\rhoDrhoPConstant
	&\triangleq
	\sqrt{\frac{\cValOne + \frac{\cValTwo}{M} + \frac{\cValFour}{M}+ \frac{\cValSix}{M}}{\cValThree+\cValFour+\cValFive}}
	\\
	\cValOne
	&\triangleq
	\sum\limits_{\sIdxTwo=0}^{N-1}	
	\frac{\ulTotPower{\sIdxTwo}{}^2\beta_{\sIdxTwo}^2}{\ulTotPower{0}{}(\ulDuration-1)}
	\;,\;
	\cValTwo
	\triangleq
	\sum\limits_{\sIdxTwo=0}^{N-1}	
	\sum\limits_{\sIdxThree=0}^{N-1}	
	\frac{\ulTotPower{\sIdxTwo}{}\ulTotPower{\sIdxThree}{}\beta_{\sIdxTwo}\beta_{\sIdxThree}}{\ulTotPower{0}{}(\ulDuration-1)}
	\;;\;
	\cValThree
	\triangleq
	\sum\limits_{\sIdxTwo=0}^{N-1}	
	\frac{\ulTotPower{\sIdxTwo}{}d_{\sIdxTwo}\ulDuration}{(\ulDuration-1)}
	\\		
	\cValFour
	&\triangleq
	\sum\limits_{\sIdxTwo=0}^{N-1}	
	\frac{\ulTotPower{\sIdxTwo}{}e_{\sIdxTwo}\ulDuration}{\ulDuration-1}				
	+
	\frac{\sigma^4}{\ulTotPower{0}{}\ulDuration}
	\;;\;
	\cValFive
	\triangleq
	\sum\limits_{\sIdxTwo\in\pilotSharingSet_0}
	\sigma^2
	\frac{\ulTotPower{\sIdxTwo}{}}{\ulTotPower{0}{}}
	\beta_{\sIdxTwo}
	\;;\;
	\cValSix
	\triangleq
	\sum\limits_{\sIdxTwo=0}^{N-1}
	\frac{\sigma^2}{\ulDuration}		
	\frac{\ulTotPower{\sIdxTwo}{}}{\ulTotPower{0}{}}
	\beta_{\sIdxTwo}\;.
	\end{align}	
	\begin{proof}
		The achievable rate in the UL as well as expressions for $ \rhoD{}{}^2_{\mathrm{opt}} $ and $ \rhoP{}{}^2_{\mathrm{opt}} $ are derived in Appendix \ref{appdx:optRhoDRhoP}.
	\end{proof}
\end{prop}

\begin{remark}
	Note that the exact expressions for achievable rate for SP when $ LK \geq \ulDuration $ have been derived earlier in \cite{verenzuela2017spectral} (c.f. Theorem~1 and Corollary~1). However, Theorem~1 in \cite{verenzuela2017spectral} underestimates the UL rate since it treats the pilot that is transmitted alongside data in each UL symbol as interference, whereas Corollary~1 over-estimates the rate since it assumes that the pilots are perfectly removed. On the other hand, in the expression for the UL rate derived in Appendix \ref{appdx:optRhoDRhoP}, we side-step this issue by multiplying the received observations with a unitary matrix that relegates all the interference resulting from transmitting pilots alongside data to a single symbol. This symbol can then be discarded since we are anyway interested in only a lower bound on the ergodic capacity. The remaining $ \ulDuration-1 $ symbols of the reference user are free of interference from the UL pilot of that user, and therefore, standard methods can be used to calculate the UL throughput in these symbols. 
\end{remark}

Substituting \eqref{eqn:rhoDUlOptimal} and \eqref{eqn:rhoPUlOptimal} into \eqref{eqn:spMse}, the expression for the NMSE becomes
\begin{align}
	\spMse_{\vIdxOne,\vIdxTwo}\bigg|_{\rhoD{}{}_{\mathrm{opt}},\rhoP{}{}_{\mathrm{opt}}}
	&=
	\frac{1}{\beta_{\vIdxOne\vIdxOne\vIdxTwo}}
	\left(
	\frac{1}{\sqrt{M}\rhoDrhoPConstant\ulDuration} 
	\sum\limits_{\sIdxOne=0}^{L-1}
	\sum\limits_{\sIdxTwo=0}^{K-1}
	\frac{\ulTotPower{\sIdxOne}{\sIdxTwo}}{\ulTotPower{\vIdxOne}{\vIdxTwo}}
	\beta_{\vIdxOne\sIdxOne\sIdxTwo}
	+
	\sum_{\cellSharingPilotSetSp{\vIdxOne}\ni\sIdxOne\neq\vIdxOne}
	\frac{\ulTotPower{\sIdxOne}{\vIdxTwo}}{\ulTotPower{\vIdxOne}{\vIdxTwo}}
	\beta_{\vIdxOne\sIdxOne\vIdxTwo}	
	\nonumber
	\right.
	\\
	&\left.
	+
	\lrc{1+{\frac{1}{\sqrt{M}\rhoDrhoPConstant}} }
	\frac{\sigma^2}{\ulDuration\ulTotPower{\vIdxOne}{\vIdxTwo}}\right) \;.
	\label{eqn:spMseOptRhoLambda}
\end{align}
Thus, with optimized values of $ \rhoD{}{}^2 $ and $ \rhoP{}{}^2 $, the component of the $ \spMse_{\vIdxOne,\vIdxTwo} $ resulting from transmitting data alongside pilots reduces proportional to the square root of the number of antenna elements. This behavior is in contrast to \eqref{eqn:cpMse}, wherein $ \cpMse_{\vIdxOne,\vIdxTwo} $ is independent of $ M $. Consequently, the reduction in the NMSE also leads to a higher DL throughput, as shown below. 

Substituting \eqref{eqn:rhoDUlOptimal} and \eqref{eqn:rhoPUlOptimal} into \eqref{eqn:spDlSinrMf}, the expression for the DL SINR becomes
\begin{align}
\spDlSinr_{\vIdxOne\vIdxTwo}
&=
\frac{\txPrecoderNormalization_{\vIdxOne\vIdxTwo}\beta_{\vIdxOne\vIdxOne\vIdxTwo}^2}
{\sum\limits_{\cellSharingPilotSetSp{\vIdxOne}\ni\sIdxOne\neq\vIdxOne}
\frac{\ulTotPower{\vIdxOne}{\vIdxTwo}}{\ulTotPower{\sIdxOne}{\vIdxTwo}}
\txPrecoderNormalization_{\sIdxOne\vIdxTwo}
\beta_{\sIdxOne\vIdxOne\vIdxTwo}^2
+
\frac{\tTermOne}{\sqrt{M}\rhoDrhoPConstant\ulDuration}
+
\frac{\tTermTwo}{M}}
\label{eqn:spDlSinrMfOptRho}
\end{align}
where
\begin{align}
	\tTermOne
	&=
	\sum\limits_{\sIdxOne=0}^{L-1}
	\sum\limits_{\sIdxTwo=0}^{K-1}			
	\txPrecoderNormalization_{\sIdxOne\sIdxTwo}
	\frac{\ulTotPower{\vIdxOne}{\vIdxTwo}}{\ulTotPower{\sIdxOne}{\sIdxTwo}}
	\beta_{\sIdxOne\vIdxOne\vIdxTwo}^2
	\\
	\tTermTwo
	&=
	\sum\limits_{\sIdxOne=0}^{L-1}
	\sum\limits_{\sIdxTwo=0}^{K-1}
	\frac{\txPrecoderNormalization_{\sIdxOne\sIdxTwo}
	\beta_{\sIdxOne\vIdxOne\vIdxTwo}}{\ulTotPower{\sIdxOne}{\sIdxTwo}}
	\lrc{
		\sum\limits_{\sIdxThree\in\cellSharingPilotSetSp{\sIdxOne}}	
		\ulTotPower{\sIdxThree}{\sIdxTwo}	
		\beta_{\sIdxOne\sIdxThree\sIdxTwo}
		+
		\lrc{1 + \frac{1}{\sqrt{M}\rhoDrhoPConstant}}
		\frac{\sigma^2}{\ulDuration}
	}
	\nonumber
	\\
	&+
		\sum\limits_{\sIdxOne=0}^{L-1}
		\sum\limits_{\sIdxTwo=0}^{L-1}
		\sum\limits_{\sIdxThree=0}^{L-1}		
		\sum\limits_{\sIdxFour=0}^{K-1}				
		\frac{
			\ulTotPower{\sIdxThree}{\sIdxFour}
			\txPrecoderNormalization_{\sIdxOne\sIdxTwo}
			\beta_{\sIdxOne\sIdxThree\sIdxFour}
			\beta_{\sIdxOne\vIdxOne\vIdxTwo}
			}{\sqrt{M}\ulTotPower{\sIdxOne}{\sIdxTwo}\rhoDrhoPConstant\ulDuration}
	+
	\frac{1}{M\dlSnr}\;.
\end{align}
 From the above expression, it can be observed that $ \tTermOne/\sqrt{M} $, which is the component of interference from users that do not share a pilot with user $ \lrc{\vIdxOne,\vIdxTwo} $, decreases proportional to the square root of $ M $. Since $ \ulDuration \gg K $, more orthogonal pilot sequences are available when SPs are employed in comparison with RP. As a result, SP can be reused over a larger number of cells, i.e., $ \pilotReuseRatioSp > \pilotReuseRatioTp $. Therefore, we have the following result.
 \begin{prop}
 	If $ \pilotReuseRatioSp > \pilotReuseRatioTp $, the ceiling of $ \spDlSinr_{\vIdxOne\vIdxTwo} $ when $ M\rightarrow\infty $ is higher than that of $ \cpDlSinr_{\vIdxOne\vIdxTwo} $.
 	\begin{proof}
 		When $ M\rightarrow\infty $ \eqref{eqn:spDlSinrMf} becomes 		
 		\begin{align}
 		\spDlSinr_{\vIdxOne\vIdxTwo}
 		&=
 		\frac{\widetilde{\txPrecoderNormalization}_{\vIdxOne\vIdxTwo}\beta_{\vIdxOne\vIdxOne\vIdxTwo}^2}
 		{\sum\limits_{\cellSharingPilotSetSp{\vIdxOne}\ni\sIdxOne\neq\vIdxOne}
 			\frac{\ulTotPower{\vIdxOne}{\vIdxTwo}}{\ulTotPower{\sIdxOne}{\vIdxTwo}}
	 			\widetilde{\txPrecoderNormalization}_{\sIdxOne\vIdxTwo}
 			\beta_{\sIdxOne\vIdxOne\vIdxTwo}^2
		}\;.
		\label{eqn:spDlSinrMfInfM}
 		\end{align}
 		Therefore, comparing \eqref{eqn:cpDlSinrMfInfM} and \eqref{eqn:spDlSinrMfInfM}, the denominator of \eqref{eqn:spDlSinrMfInfM} is smaller than that of \eqref{eqn:cpDlSinrMfInfM} when $ \pilotReuseRatioSp > \pilotReuseRatioTp $. Here $ \widetilde{\txPrecoderNormalization}_{\vIdxOne\vIdxTwo} = \lim_{M\rightarrow\infty} M \txPrecoderNormalization_{\vIdxOne\vIdxTwo} $ and is independent of $ M $.
 	\end{proof}
 	\label{prop:SpVsRp}
 \end{prop}
 

%% file: texFiles/staggeredPilot.tex
 \section{Staggered Pilots as a Particular Case of Superimposed Pilots}
 \label{sec:staggeredPilots}
 
 When transmitting staggered pilots \cite{kong2016Channel,mahyiddin2015performance}, the users in each cell stagger their pilot transmissions so that the users of no two cells within the $ \pilotReuseRatioSp $ cells, which share the $ \ulDuration $ UL pilots, are transmitting UL pilots simultaneously, i.e., if the users in a particular cell are transmitting UL pilots, the users in the remaining $ \pilotReuseRatioSp-1 $ cells transmit data. Let $ \mbf{Y}_{\sIdxThree}$ be the observation at BS $ \vIdxOne $ when the users in the $ \sIdxThree\rth $ cell (where $ 0\leq\sIdxThree\leq\pilotReuseRatioSp-1 $) transmit UL pilots. Note that the index $ \vIdxOne $ has been dropped from $ \mbf{Y}_{\sIdxThree} $ for the sake of simplicity of notation. Then, $ \mbf{Y}_{\sIdxThree} \in \mathbb{C}^{M\times\tau} $ can be written as
 \begin{align}
 	\mbf{Y}_{\sIdxThree}
 	&\triangleq
 	\sum\limits_{\sIdxOne\in\cellSharingPilotSetSp{\sIdxThree}}
 	\sum\limits_{\sIdxTwo} 	
 	\sqrt{\ulTotPower{\sIdxOne}{\sIdxTwo}\stagPPilotPower}
 	\mbf{h}_{\vIdxOne\sIdxOne\sIdxTwo}
 	\cpPilot_{\sIdxThree\sIdxTwo}^T
 	+
 	\sum\limits_{\sIdxOne\notin\cellSharingPilotSetSp{\sIdxThree}}
 	\sum\limits_{\sIdxTwo}
 	\sqrt{\ulTotPower{\sIdxOne}{\sIdxTwo}\stagPDataPower}
 	\mbf{h}_{\vIdxOne\sIdxOne\sIdxTwo}
 	\lrc{\ulTxData_{\sIdxOne\sIdxTwo}^{\sIdxThree}}^T
 	+
 	\mbf{W}_{\sIdxThree}
 	\label{eqn:stagPilotUl}
 \end{align}
 where $ \cpPilot_{\sIdxThree,\sIdxTwo} \; \forall \sIdxThree,\sIdxTwo  $ are the orthogonal pilot sequences described in Subsection \ref{sec:pilotContaminationDownlinkTp}, $ \stagPPilotPower $ and $ \stagPDataPower $ are the powers at which the pilots and the data, respectively, are transmitted, and $ \ulTxData_{\sIdxOne,\sIdxTwo}^{\sIdxThree} \in \mathbb{C}^{\tau} $ is the vector of data symbols transmitted by user $ \lrc{\sIdxOne,\sIdxTwo} $ in the $ \sIdxThree\rth $ block. 
 We then have the following proposition.
 \begin{prop}
   The UL in a system that employs staggered pilots in \eqref{eqn:stagPilotUl} is a particular case of superimposed pilots if $ \stagPPilotPower = \ulTotPower{}{}\rhoP{}{}^2\ulDuration/\tau $, $ \stagPDataPower = \ulTotPower{}{}\rhoD{}{}^2 $ and $ \mbf{P} = \sqrt{\frac{\ulDuration}{\tau}} \mathrm{blkdiag}\lrf{\pmb{\Phi}_{0},\ldots,\pmb{\Phi}_{L-1}} $.
   \label{prop:staggeredPilot}
 \end{prop}
An important conclusion of Proposition \ref{prop:staggeredPilot} is that staggered pilots are capable of achieving the downlink throughput of SP while maintaining the UL spectral efficiency of RP. Indeed, utilizing the same approach used to derive \eqref{eqn:spDlSinrMf}, a lower bound on the DL ergodic capacity when the channel estimate obtained from staggered pilots is employed in a MF precoder can be obtained as
\begin{align}
	\stagDlRate_{\vIdxOne\vIdxTwo} = \frac{\dlDuration}{\coherenceTime} \log_2\lrc{1 + \stagDlSinr_{\vIdxOne\vIdxTwo}}
\end{align}
where
\begin{align}
	&\stagDlSinr_{\vIdxOne\vIdxTwo}
	=
	\txPrecoderNormalization_{\vIdxOne\vIdxTwo}
	\beta_{\vIdxOne\vIdxOne\vIdxTwo}^2
	\!\!
	\left(		
		\sum\limits_{\cellSharingPilotSetSp{\vIdxOne}\ni\sIdxOne\neq\vIdxOne}
		\!\!
		\txPrecoderNormalization_{\sIdxOne\vIdxTwo}
		\frac{\ulTotPower{\vIdxOne}{\vIdxTwo}}{\ulTotPower{\sIdxOne}{\vIdxTwo}}
		\beta_{\vIdxOne\sIdxOne\vIdxTwo}^2
		+
		\frac{1}{M}
		\sum\limits_{\sIdxOne=0}^{L-1}
		\sum\limits_{\sIdxTwo=0}^{K-1}		
		\frac{\txPrecoderNormalization_{\sIdxOne\sIdxTwo}
		\beta_{\sIdxOne\vIdxOne\vIdxTwo}}{\ulTotPower{\sIdxOne}{\sIdxTwo}}
		\lrc{
		\sum\limits_{\sIdxThree\in\cellSharingPilotSetSp{\sIdxOne}}
		\ulTotPower{\sIdxThree}{\sIdxTwo}
		\beta_{\sIdxOne\sIdxThree\sIdxTwo}
		+
		\frac{\sigma^2}{\stagPPilotPower\tau}
		}
		\right.
		\nonumber
		\\
		&\left.
		+
		\frac{\stagPDataPower}{\tau\stagPPilotPower}
		\sum\limits_{\sIdxOne=0}^{L-1}
		\sum\limits_{\sIdxTwo=0}^{K-1}
		\frac{\txPrecoderNormalization_{\sIdxOne\sIdxTwo}}{\ulTotPower{\sIdxOne}{\sIdxTwo}}
		\lrc{
			\ulTotPower{\vIdxOne}{\vIdxTwo}
			\beta_{\sIdxOne\vIdxOne\vIdxTwo}^2
			\indicator{\vIdxOne\notin\cellSharingPilotSetSp{\sIdxOne}}
			+
			\frac{1}{M}
			\sum\limits_{\sIdxThree\notin\cellSharingPilotSetSp{\sIdxOne}}
			\sum\limits_{\sIdxFour=0}^{K-1}
			\ulTotPower{\sIdxThree}{\sIdxFour}
			\beta_{\sIdxOne\vIdxOne\vIdxTwo}\beta_{\sIdxOne\sIdxThree\sIdxFour}
			}		
		+
		\frac{1}{M^2\dlSnr}
	\right)^{-1}
	\label{eqn:stagDlSinrMf}
\end{align} 
and	
\begin{align}
	\txPrecoderNormalization_{\sIdxOne\sIdxTwo}
	=
	\frac{\dlTxPower_{\sIdxOne\sIdxTwo}}{\expectation\lrf{\|\txPrecoder_{\sIdxOne\sIdxTwo}\|^2}}
	=
	\frac{\dlTxPower_{\sIdxOne\sIdxTwo}}{M}
	\lrc{
			\sum\limits_{\sIdxThree\in\cellSharingPilotSetSp{\sIdxOne}}
			\frac{\ulTotPower{\sIdxThree}{\sIdxTwo}}{\ulTotPower{\sIdxOne}{\sIdxTwo}}
			\beta_{\sIdxOne\sIdxThree\sIdxTwo}
			+
			\frac{\stagPDataPower}{\tau\stagPPilotPower}
			\sum\limits_{\sIdxThree\notin\cellSharingPilotSetSp{\sIdxOne}}
			\sum\limits_{\sIdxFour=0}^{K-1}
			\frac{\ulTotPower{\sIdxThree}{\sIdxFour}}{\ulTotPower{\sIdxOne}{\sIdxTwo}}
			\beta_{\sIdxOne\sIdxThree\sIdxFour}
			+
			\frac{\sigma^2}{\tau\stagPPilotPower\ulTotPower{\sIdxOne}{\sIdxTwo}}
		}^{-1}\;.
\end{align}
Therefore, staggered pilots can achieve the DL performance of RP with a reuse factor $ \pilotReuseRatioSp $ with an overhead equivalent to that of RP with pilot-reuse factor $ \pilotReuseRatioTp $. As a result, similar to Proposition~\ref{prop:SpVsRp}, we have
\begin{prop}
	If $ \pilotReuseRatioSp > \pilotReuseRatioTp $, the ceiling of $ \stagDlSinr_{\vIdxOne\vIdxTwo} $ when $ M\rightarrow\infty $ is higher than that of $ \cpDlSinr_{\vIdxOne\vIdxTwo} $.
\end{prop}

The concept described in this section can be further demonstrated by a simple example. Consider a system with two users A and B. Without loss of generality, it is assumed that the large-scale path-loss between the BS and users A and B are unity. In the UL phase, let user B transmit data with power $ \rhoD{}{}^2\geq 0 $ when user A transmits its pilot at unit power. In the DL phase, user B receives interference at a power $ \rhoD{}{}^2 $ from the DL transmission to user A. Thus, increasing the number of antennas $ M $ at the BS increases the array gain at the BS, allowing for user B to transmit with a smaller power $ \rhoD{}{}^2 $, thereby reducing the interference it sees in the DL. 

%% file: texFiles/hybridSystem.tex
\section{Extension of Hybrid System to DL}
\label{sec:hybridSystem}
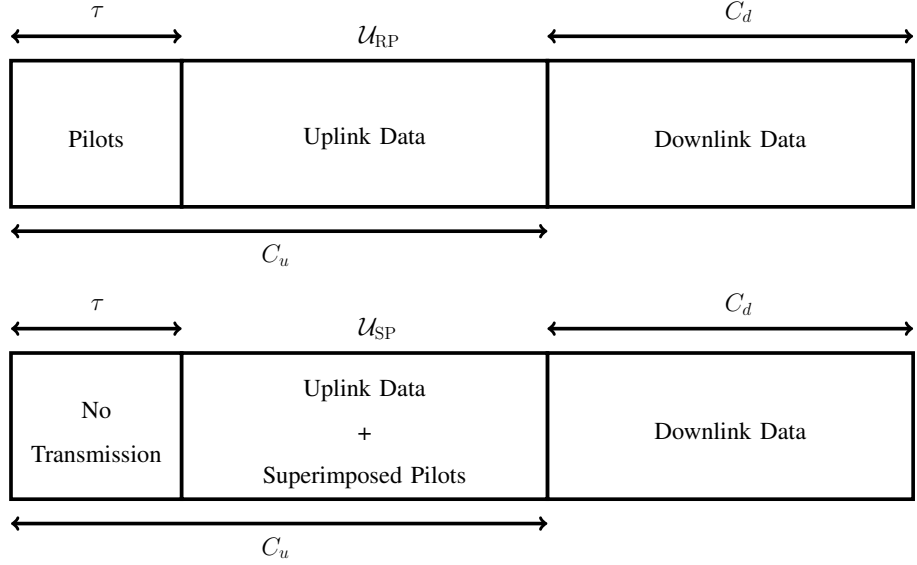
\begin{figure}
	\centering
	\iftoggle{oneColumn}
	{
		\resizebox{0.75\textwidth}{!}{\input{figs/theorem1fig.tex}}	
	}
	{
		\resizebox{0.5\textwidth}{!}{\input{figs/theorem1fig.tex}}	
	}
	\caption{Frame structure of a hybrid system with users employing RP and SP pilots.}
	\label{fig:hybridSystem}
\end{figure}
Using the DL SINR analysis in Section \ref{sec:pilotContaminationDownlink}, we extend the hybrid system in \cite{upadhya2016superimposed} to include the DL as well. The hybrid system consists of two sets of users $ \cpUserSet $ and $ \spUserSet $ that transmit RP and SP, respectively. As shown in Fig. \ref{fig:hybridSystem}, users in $ \cpUserSet $ obtain channel estimates in the UL using RPs, transmitted over $ \tau $ symbols, and use these estimates to detect data using a spatial filter. However, users in $ \spUserSet $ maintain radio silence during the pilot training phase of the users in $ \cpUserSet $, i.e., for $ \tau $ symbols in the frame, and transmit orthogonal pilots superimposed with data during the UL data phase of $ \ulDuration-\tau $ symbols. By this construction, users in $ \spUserSet $ effectively transmit the zero vector for the $ \tau $ training symbols and since the zero vector is orthogonal to all other vectors, the users in $ \spUserSet $ can be viewed as having orthogonal pilots, thus not affecting the performance of any user in $ \cpUserSet $.

Employing the channel estimates obtained from $ \cpUserSet $ and $ \spUserSet $ in a MF precoder and combiner, the SINR in the UL (see \cite{upadhya2016superimposed}) and DL from users in $ \cpUserSet $ and $ \spUserSet $ when $ M\gg K $ \footnote{ Since this section is devoted to designing a suboptimal algorithm to partition users into those that use RP and SP, we rely on approximate asymptotic expressions \eqref{eqn:hybridCpDlSinr} -- \eqref{eqn:hybridSpDlSinr}, for the sake of simplicity, since the problem is anyway solved sub-optimally.} can be obtained as
\begingroup
\allowdisplaybreaks
\begin{align}
	\cpUlSinr_{\vIdxOne\vIdxTwo}
	&=
	\frac
	{
		\beta_{\vIdxOne\vIdxOne\vIdxTwo}^2
	}
	{
		\!\!\!\!\!\!\!
		\sum\limits_{\substack{\sIdxOne\neq\vIdxOne\\\lrc{\sIdxOne,\vIdxTwo}\in\cpUserSet}}
		\!\!\!\!\!\!\!
		\beta_{\vIdxOne\sIdxOne\vIdxTwo}^2
	}
	\label{eqn:hybridCpUlSinr}
	\\
	\cpDlSinr_{\vIdxOne\vIdxTwo}
	&=
	\frac
	{
		\widetilde{\txPrecoderNormalization}_{\vIdxOne\vIdxTwo}
		\beta_{\vIdxOne\vIdxOne\vIdxTwo}^2
	}
	{
		\!\!\!\!
		\sum\limits_{\substack{\sIdxOne\neq\vIdxOne\\\lrc{\sIdxOne\vIdxTwo}\in\cpUserSet}}
		\!\!\!\!
		\widetilde{\txPrecoderNormalization}_{\sIdxOne\vIdxTwo}
		\beta_{\sIdxOne\vIdxOne\vIdxTwo}^2
	}
	\label{eqn:hybridCpDlSinr}	
	\\
	\sinrUlNonIterative_{\vIdxOne\vIdxTwo}
	&\approx 
	\frac{\beta_{\vIdxOne\vIdxOne\vIdxTwo}^2}
	{
		\frac{1}{\lrc{\ulDuration-\tau}\rhoP{}{}^2}
		\underset{{\lrc{\sIdxOne,\sIdxTwo}\in\spUserSet}}
		{
			\sum\limits_{\sIdxOne}
			\sum\limits_{\sIdxTwo}
		}
		\beta_{\vIdxOne\sIdxOne\sIdxTwo}^2
	}
	\label{eqn:hybridSpUlSinr}
	\\
	\spDlSinr_{\vIdxOne\vIdxTwo}
	&\approx
	\frac{\widetilde{\txPrecoderNormalization}_{\vIdxOne\vIdxTwo}\beta_{\vIdxOne\vIdxOne\vIdxTwo}^2}
	{	
		\frac{\rhoD{}{}^2}{\lrc{\ulDuration-\tau}\rhoP{}{}^2}	
		\underset{\lrc{\sIdxOne,\sIdxTwo}\in\spUserSet}{
			\sum\limits_{\sIdxOne}
			\sum\limits_{\sIdxTwo}
		}
		\widetilde{\txPrecoderNormalization}_{\sIdxOne\vIdxTwo}
		\beta_{\sIdxOne\vIdxOne\vIdxTwo}^2		
	}
	\label{eqn:hybridSpDlSinr}
\end{align}
\endgroup
where the approximations in \eqref{eqn:hybridSpUlSinr}, and \eqref{eqn:hybridSpDlSinr} reflect on the assumption that the users in $ \cpUserSet $ and $ \spUserSet $ do not interfere with each other. This assumption is valid if the UL transmission power of the users in $ \cpUserSet $ is significantly smaller than those in $ \spUserSet $. This assumption is made for the sake of simplicity and clarity only. In the absence of this assumption, the BS will have to estimate and remove the interference from the users in $ \cpUserSet $ before estimating the channel vectors of the users in $ \spUserSet $.  In addition, for the sake of simplicity, we assume that $ \pilotReuseRatioTp = 1 $ and that the interference from the cells other than the ones adjacent to the reference cell are negligible. Furthermore, the UL transmit powers $ \ulTotPower{\sIdxOne}{\sIdxTwo} $ are subsumed into the coefficients $ \beta_{\vIdxOne\sIdxOne\sIdxTwo} $.

In \cite{upadhya2016superimposed}, the objective of the hybrid system design has been defined as to partition the users into disjoint sets $ \cpUserSet $ and $ \spUserSet $ by minimizing the overall UL interference. Using \eqref{eqn:hybridCpDlSinr} and \eqref{eqn:hybridSpUlSinr}, we extend here the objective to jointly minimize the UL and DL interference. 

Let $ \hybridCpUlIci_{\vIdxOne\vIdxTwo} $ or $ \hybridCpDlIci_{\vIdxOne\vIdxTwo} $, respectively, be the contributions of user $ \lrc{\vIdxOne,\vIdxTwo} $ to the total UL and DL inter/intra-cell interference power when assigned to $ \cpUserSet $. Similarly, let $ \hybridSpUlIci_{\vIdxOne\vIdxTwo} $ or $ \hybridSpDlIci_{\vIdxOne\vIdxTwo} $, respectively, be the contributions of user $ \lrc{\vIdxOne,\vIdxTwo} $ to the total DL inter/intra-cell interference power when assigned to $ \spUserSet $. If users $ \lrc{\vIdxOne,\vIdxTwo} $ and $ \lrc{\sIdxOne,\sIdxTwo} $ are members of $ \cpUserSet  $, then from the denominator of \eqref{eqn:hybridCpUlSinr}, the amount of interference that user $ \lrc{\vIdxOne,\vIdxTwo} $ causes to user $ \lrc{\sIdxOne,\sIdxTwo} $ in the UL is $ \beta_{\sIdxOne\vIdxOne\sIdxTwo}^2 \delta_{\vIdxTwo,\sIdxTwo} $. Similarly, from \eqref{eqn:hybridCpDlSinr}, the amount of interference that user $ \lrc{\vIdxOne,\vIdxTwo} $ causes to user $ \lrc{\sIdxOne,\sIdxTwo} $ in the DL is \mbox{$ \beta_{\sIdxThree\sIdxOne\sIdxTwo}^2 \delta_{\vIdxOne,\sIdxOne}\delta_{\vIdxTwo,\sIdxTwo}, \; \forall \sIdxThree\neq\sIdxOne,\; \sIdxThree\in\mathcal{L}_{\vIdxOne}\lrc{r} ,\lrc{\sIdxThree,\sIdxTwo}\in\cpUserSet$}. Likewise, from \eqref{eqn:hybridSpUlSinr} and \eqref{eqn:hybridSpDlSinr}, if both users are members of $ \spUserSet $ then the amount of interference that user $ \lrc{\vIdxOne,\vIdxTwo} $ causes to user $ \lrc{\sIdxOne,\sIdxTwo} $ in the UL and DL is $ \beta_{\sIdxOne\vIdxOne\vIdxTwo}^2/\lrc{\lrc{\ulDuration-\tau}\rhoP{}{}^2} $ and $ \rhoD{}{}^2\beta_{\sIdxThree\vIdxOne\vIdxTwo}^2/\lrc{\lrc{\ulDuration-\tau}\rhoP{}{}^2},\;\forall \sIdxThree\neq\vIdxOne,\;\sIdxThree=0,\ldots,L-1 $, respectively. Therefore, $ \hybridCpUlIci_{\vIdxOne\vIdxTwo} $, $ \hybridCpDlIci_{\vIdxOne\vIdxTwo} $, $ \hybridSpUlIci_{\vIdxOne\vIdxTwo} $, and $ \hybridSpDlIci_{\vIdxOne\vIdxTwo} $ can be obtained as
\begingroup
\allowdisplaybreaks
\begin{align}
	\hybridCpUlIci_{\vIdxOne\vIdxTwo} 
	&=
	\!\!\!\!
	\underset{\substack{\sIdxOne\in\mathcal{L}_{\vIdxOne}(\pilotReuseRatio)\\\lrc{\sIdxOne,\sIdxTwo}\in\cpUserSet}}{
	\sum\limits_{\substack{\sIdxOne\neq\vIdxOne}}
	\sum\limits_{\sIdxTwo}
	}
	\beta_{\sIdxOne\vIdxOne\sIdxTwo}^2
	\delta_{\vIdxTwo,\sIdxTwo}
	=
	\!\!\!\!
	\sum\limits_{\substack{\sIdxOne\neq\vIdxOne\\\sIdxOne\in\mathcal{L}_{\vIdxOne}(\pilotReuseRatio)\\\lrc{\sIdxOne,\vIdxTwo}\in\cpUserSet}}
	\!\!\!\!\!\!\!
	\beta_{\sIdxOne\vIdxOne\vIdxTwo}^2
	\\
	\hybridCpDlIci_{\vIdxOne,\vIdxTwo}
	&=	
	\mathop{
		\sum\limits_{\sIdxThree\neq\vIdxOne}
		\sum\limits_{\substack{\sIdxOne}}		
		\sum\limits_{\sIdxTwo}
	}_{\substack{\sIdxThree,\sIdxOne\in\mathcal{L}_{\vIdxOne}(\pilotReuseRatio)\\\lrc{\sIdxThree,\sIdxTwo}\in\cpUserSet}}
	\beta_{\sIdxThree\sIdxOne\sIdxTwo}^2 
	\delta_{\vIdxOne,\sIdxOne}
	\delta_{\vIdxTwo,\sIdxTwo}
	=
	\!\!\!\!
	\sum\limits_{\substack{\sIdxThree\neq\vIdxOne\\\sIdxThree\in\mathcal{L}_{\vIdxOne}(r)\\\lrc{\sIdxThree,\vIdxTwo}\in\cpUserSet}}
	\!\!\!\!
	\beta_{\sIdxThree\vIdxOne\vIdxTwo}^2 
	\\
	\hybridSpUlIci_{\vIdxOne\vIdxTwo} 
	&= 
	\frac{1}{\lrc{\ulDuration-\tau}\rhoP{}{}^2}
	\underset{{\lrc{\sIdxOne,\sIdxTwo}\in\spUserSet}}
	{
		\sum\limits_{\sIdxOne}
		\sum\limits_{\sIdxTwo}
	}
	\beta_{\sIdxOne\vIdxOne\vIdxTwo}^2
	\\
	\hybridSpDlIci_{\vIdxOne\vIdxTwo} 
	&= 
	\frac{\rhoD{}{}^2}{\lrc{\ulDuration-\tau}\rhoP{}{}^2}
	\underset{{\lrc{\sIdxOne,\sIdxTwo}\in\spUserSet}}
	{
		\sum\limits_{\sIdxOne}
		\sum\limits_{\sIdxTwo}
	}
	\beta_{\sIdxOne\vIdxOne\vIdxTwo}^2
	=
	\rhoD{}{}^2 \hybridSpUlIci_{\vIdxOne\vIdxTwo} 	\;.	
\end{align}
\endgroup

Let $ \hybridUlCostFactor > 0 $ and $ \hybridDlCostFactor > 0 $ be the weights for the interference powers in the UL and DL, respectively, such that \mbox{$ \hybridUlCostFactor+\hybridDlCostFactor = 1 $}. Then, the total cost due to inter/intra-cell interference can be expressed as
\begin{align}
	\hybridTotalIci\lrc{\cpUserSet,\spUserSet}
	&= 
	\sum\limits_{\sIdxOne=0}^{L-1}
	\sum\limits_{\sIdxTwo=0}^{K-1}
	\left(
	\hybridCpCost_{\sIdxOne\sIdxTwo}
	\indicator{\lrc{\sIdxOne,\sIdxTwo}\in\cpUserSet}
	+
	\hybridSpCost_{\sIdxOne\sIdxTwo} 
	\indicator{\lrc{\sIdxOne,\sIdxTwo}\in\spUserSet}	
	\right)
	\label{eqn:hybridTotalIci}
\end{align}
where $ \hybridCpCost_{\sIdxOne\sIdxTwo} $ and $ \hybridSpCost_{\sIdxOne\sIdxTwo} $ are the costs incurred when user $ \lrc{\sIdxOne,\sIdxTwo} $ is assigned to $ \cpUserSet $ and $ \spUserSet $, respectively, and are defined as
\begin{align}
	\hybridCpCost_{\sIdxOne\sIdxTwo} 
	&\triangleq
		\hybridUlCostFactor
		\hybridCpUlIci_{\sIdxOne\sIdxTwo}
		+
		\hybridDlCostFactor
		\hybridCpDlIci_{\sIdxOne\sIdxTwo}
	\\
	\hybridSpCost_{\sIdxOne\sIdxTwo} 
	&\triangleq
	\hybridUlCostFactor
	\hybridSpUlIci_{\sIdxOne\sIdxTwo}
	+
	\hybridDlCostFactor
	\hybridSpDlIci_{\sIdxOne\sIdxTwo}\;.
\end{align}%
Minimizing \eqref{eqn:hybridTotalIci} over the possible choices of $ \cpUserSet $ and $ \spUserSet 	$, the optimal sets $ \cpUserSet $ and $ \spUserSet $ can be obtained as the solution of the following optimization problem
\begin{align}
	\lrc{\cpUserSet,\spUserSet }
	=
	&\arg \min_{\substack{\cpUserSet\subseteq\hybridCompleteUserSet\\\spUserSet\subseteq\hybridCompleteUserSet}} \hybridTotalIci\lrc{\cpUserSet,\spUserSet} 
	\nonumber
	\\ 
	&\text{subject to}\quad \cpUserSet\cup\spUserSet=\hybridCompleteUserSet 
	\nonumber\\
	&\qquad\qquad\quad\cpUserSet\cap\spUserSet=\varnothing
	\label{eqn:hybridOptimizationProblem}
\end{align}
where $ \hybridCompleteUserSet $ is the set of all users in the $ L $ cells. However, obtaining the solution to the optimization problem in \eqref{eqn:hybridOptimizationProblem} is combinatorial in nature with $ 2^{\cardinality{\hybridCompleteUserSet}} $ possible choices for $ \cpUserSet $ and $ \spUserSet $. A simple greedy algorithm to partition the users by minimizing only the overall UL interference has been devised in \cite{upadhya2016superimposed}, and it can be straightforwardly extended to jointly minimize both the UL and DL interference powers.

%% file: figs/theorem1fig.tex
\ifx\du\undefined
  \newlength{\du}
\fi
\iftoggle{oneColumn}
{
	\setlength{\du}{20\unitlength}
}
{
	\setlength{\du}{15\unitlength}
}
\begin{tikzpicture}
\pgftransformxscale{1.000000}
\pgftransformyscale{-1.000000}
\definecolor{dialinecolor}{rgb}{0.000000, 0.000000, 0.000000}
\pgfsetstrokecolor{dialinecolor}
\definecolor{dialinecolor}{rgb}{1.000000, 1.000000, 1.000000}
\pgfsetfillcolor{dialinecolor}
\definecolor{dialinecolor}{rgb}{1.000000, 1.000000, 1.000000}
\pgfsetfillcolor{dialinecolor}
\fill (15.000000\du,10.000000\du)--(15.000000\du,16.000000\du)--(22.000000\du,16.000000\du)--(22.000000\du,10.000000\du)--cycle;
\pgfsetlinewidth{0.150000\du}
\pgfsetdash{}{0pt}
\pgfsetdash{}{0pt}
\pgfsetmiterjoin
\definecolor{dialinecolor}{rgb}{0.000000, 0.000000, 0.000000}
\pgfsetstrokecolor{dialinecolor}
\draw (15.000000\du,10.000000\du)--(15.000000\du,16.000000\du)--(22.000000\du,16.000000\du)--(22.000000\du,10.000000\du)--cycle;
\definecolor{dialinecolor}{rgb}{0.000000, 0.000000, 0.000000}
\pgfsetstrokecolor{dialinecolor}
\node at (18.500000\du,13.195000\du){\LARGE Pilots};
\definecolor{dialinecolor}{rgb}{1.000000, 1.000000, 1.000000}
\pgfsetfillcolor{dialinecolor}
\fill (37.000000\du,10.000000\du)--(37.000000\du,16.000000\du)--(52.000000\du,16.000000\du)--(52.000000\du,10.000000\du)--cycle;
\pgfsetlinewidth{0.150000\du}
\pgfsetdash{}{0pt}
\pgfsetdash{}{0pt}
\pgfsetmiterjoin
\definecolor{dialinecolor}{rgb}{0.000000, 0.000000, 0.000000}
\pgfsetstrokecolor{dialinecolor}
\draw (37.000000\du,10.000000\du)--(37.000000\du,16.000000\du)--(52.000000\du,16.000000\du)--(52.000000\du,10.000000\du)--cycle;
\definecolor{dialinecolor}{rgb}{0.000000, 0.000000, 0.000000}
\pgfsetstrokecolor{dialinecolor}
\node at (44.500000\du,13.195000\du){\LARGE Downlink Data};
\definecolor{dialinecolor}{rgb}{1.000000, 1.000000, 1.000000}
\pgfsetfillcolor{dialinecolor}
\fill (22.000000\du,10.000000\du)--(22.000000\du,16.000000\du)--(37.000000\du,16.000000\du)--(37.000000\du,10.000000\du)--cycle;
\pgfsetlinewidth{0.150000\du}
\pgfsetdash{}{0pt}
\pgfsetdash{}{0pt}
\pgfsetmiterjoin
\definecolor{dialinecolor}{rgb}{0.000000, 0.000000, 0.000000}
\pgfsetstrokecolor{dialinecolor}
\draw (22.000000\du,10.000000\du)--(22.000000\du,16.000000\du)--(37.000000\du,16.000000\du)--(37.000000\du,10.000000\du)--cycle;
\definecolor{dialinecolor}{rgb}{0.000000, 0.000000, 0.000000}
\pgfsetstrokecolor{dialinecolor}
\node at (29.500000\du,13.195000\du){\LARGE Uplink Data};
\definecolor{dialinecolor}{rgb}{1.000000, 1.000000, 1.000000}
\pgfsetfillcolor{dialinecolor}
\fill (15.000000\du,22.000000\du)--(15.000000\du,28.000000\du)--(22.000000\du,28.000000\du)--(22.000000\du,22.000000\du)--cycle;
\pgfsetlinewidth{0.150000\du}
\pgfsetdash{}{0pt}
\pgfsetdash{}{0pt}
\pgfsetmiterjoin
\definecolor{dialinecolor}{rgb}{0.000000, 0.000000, 0.000000}
\pgfsetstrokecolor{dialinecolor}
\draw (15.000000\du,22.000000\du)--(15.000000\du,28.000000\du)--(22.000000\du,28.000000\du)--(22.000000\du,22.000000\du)--cycle;
\definecolor{dialinecolor}{rgb}{0.000000, 0.000000, 0.000000}
\pgfsetstrokecolor{dialinecolor}
\node at (18.500000\du,25.195000\du){\LARGE \begin{tabular}{c}No \\ Transmission\end{tabular}};
\definecolor{dialinecolor}{rgb}{1.000000, 1.000000, 1.000000}
\pgfsetfillcolor{dialinecolor}
\fill (37.000000\du,22.000000\du)--(37.000000\du,28.000000\du)--(52.000000\du,28.000000\du)--(52.000000\du,22.000000\du)--cycle;
\pgfsetlinewidth{0.150000\du}
\pgfsetdash{}{0pt}
\pgfsetdash{}{0pt}
\pgfsetmiterjoin
\definecolor{dialinecolor}{rgb}{0.000000, 0.000000, 0.000000}
\pgfsetstrokecolor{dialinecolor}
\draw (37.000000\du,22.000000\du)--(37.000000\du,28.000000\du)--(52.000000\du,28.000000\du)--(52.000000\du,22.000000\du)--cycle;
\definecolor{dialinecolor}{rgb}{0.000000, 0.000000, 0.000000}
\pgfsetstrokecolor{dialinecolor}
\node at (44.500000\du,25.195000\du){\LARGE Downlink Data};
\definecolor{dialinecolor}{rgb}{1.000000, 1.000000, 1.000000}
\pgfsetfillcolor{dialinecolor}
\fill (22.000000\du,22.000000\du)--(22.000000\du,28.000000\du)--(37.000000\du,28.000000\du)--(37.000000\du,22.000000\du)--cycle;
\pgfsetlinewidth{0.150000\du}
\pgfsetdash{}{0pt}
\pgfsetdash{}{0pt}
\pgfsetmiterjoin
\definecolor{dialinecolor}{rgb}{0.000000, 0.000000, 0.000000}
\pgfsetstrokecolor{dialinecolor}
\draw (22.000000\du,22.000000\du)--(22.000000\du,28.000000\du)--(37.000000\du,28.000000\du)--(37.000000\du,22.000000\du)--cycle;
\definecolor{dialinecolor}{rgb}{0.000000, 0.000000, 0.000000}
\pgfsetstrokecolor{dialinecolor}
\node at (29.500000\du,25.195000\du){\LARGE\begin{tabular}{c}Uplink Data \\+\\ Superimposed Pilots\end{tabular}};
\definecolor{dialinecolor}{rgb}{0.000000, 0.000000, 0.000000}
\pgfsetstrokecolor{dialinecolor}
\node[anchor=west] at (18.500000\du,13.000000\du){};
\definecolor{dialinecolor}{rgb}{0.000000, 0.000000, 0.000000}
\pgfsetstrokecolor{dialinecolor}
\node[anchor=west] at (29.500000\du,13.000000\du){};
\pgfsetlinewidth{0.150000\du}
\pgfsetdash{}{0pt}
\pgfsetdash{}{0pt}
\pgfsetbuttcap
{
\definecolor{dialinecolor}{rgb}{0.000000, 0.000000, 0.000000}
\pgfsetfillcolor{dialinecolor}
\pgfsetarrowsstart{to}
\pgfsetarrowsend{to}
\definecolor{dialinecolor}{rgb}{0.000000, 0.000000, 0.000000}
\pgfsetstrokecolor{dialinecolor}
\draw (15.000000\du,9.000000\du)--(22.000000\du,9.000000\du);
}
\definecolor{dialinecolor}{rgb}{0.000000, 0.000000, 0.000000}
\pgfsetstrokecolor{dialinecolor}
\node[anchor=west] at (18.000000\du,8.000000\du){\LARGE$\tau$};
\pgfsetlinewidth{0.150000\du}
\pgfsetdash{}{0pt}
\pgfsetdash{}{0pt}
\pgfsetbuttcap
{
\definecolor{dialinecolor}{rgb}{0.000000, 0.000000, 0.000000}
\pgfsetfillcolor{dialinecolor}
\pgfsetarrowsstart{to}
\pgfsetarrowsend{to}
\definecolor{dialinecolor}{rgb}{0.000000, 0.000000, 0.000000}
\pgfsetstrokecolor{dialinecolor}
\draw (15.000000\du,21.000000\du)--(22.000000\du,21.000000\du);
}
\definecolor{dialinecolor}{rgb}{0.000000, 0.000000, 0.000000}
\pgfsetstrokecolor{dialinecolor}
\node[anchor=west] at (18.000000\du,20.000000\du){\LARGE$\tau$};
\definecolor{dialinecolor}{rgb}{0.000000, 0.000000, 0.000000}
\pgfsetstrokecolor{dialinecolor}
\node[anchor=west] at (29.000000\du,9.000000\du){\LARGE$\cpUserSet$};
\definecolor{dialinecolor}{rgb}{0.000000, 0.000000, 0.000000}
\pgfsetstrokecolor{dialinecolor}
\node[anchor=west] at (29.000000\du,21.000000\du){\LARGE$\spUserSet$};
\pgfsetlinewidth{0.150000\du}
\pgfsetdash{}{0pt}
\pgfsetdash{}{0pt}
\pgfsetbuttcap
{
\definecolor{dialinecolor}{rgb}{0.000000, 0.000000, 0.000000}
\pgfsetfillcolor{dialinecolor}
\pgfsetarrowsstart{to}
\pgfsetarrowsend{to}
\definecolor{dialinecolor}{rgb}{0.000000, 0.000000, 0.000000}
\pgfsetstrokecolor{dialinecolor}
\draw (15.000000\du,29.000000\du)--(37.000000\du,29.000000\du);
}
\definecolor{dialinecolor}{rgb}{0.000000, 0.000000, 0.000000}
\pgfsetstrokecolor{dialinecolor}
\node[anchor=west] at (25.000000\du,30.000000\du){\LARGE$C_u$};
\pgfsetlinewidth{0.150000\du}
\pgfsetdash{}{0pt}
\pgfsetdash{}{0pt}
\pgfsetbuttcap
{
\definecolor{dialinecolor}{rgb}{0.000000, 0.000000, 0.000000}
\pgfsetfillcolor{dialinecolor}
\pgfsetarrowsstart{to}
\pgfsetarrowsend{to}
\definecolor{dialinecolor}{rgb}{0.000000, 0.000000, 0.000000}
\pgfsetstrokecolor{dialinecolor}
\draw (15.000000\du,17.000000\du)--(37.000000\du,17.000000\du);
}
\definecolor{dialinecolor}{rgb}{0.000000, 0.000000, 0.000000}
\pgfsetstrokecolor{dialinecolor}
\node[anchor=west] at (25.000000\du,18.000000\du){\LARGE$C_u$};
\pgfsetlinewidth{0.150000\du}
\pgfsetdash{}{0pt}
\pgfsetdash{}{0pt}
\pgfsetbuttcap
{
\definecolor{dialinecolor}{rgb}{0.000000, 0.000000, 0.000000}
\pgfsetfillcolor{dialinecolor}
\pgfsetarrowsstart{to}
\pgfsetarrowsend{to}
\definecolor{dialinecolor}{rgb}{0.000000, 0.000000, 0.000000}
\pgfsetstrokecolor{dialinecolor}
\draw (37.000000\du,9.000000\du)--(52.000000\du,9.000000\du);
}
\definecolor{dialinecolor}{rgb}{0.000000, 0.000000, 0.000000}
\pgfsetstrokecolor{dialinecolor}
\node[anchor=west] at (44.000000\du,8.000000\du){\LARGE$C_d$};
\pgfsetlinewidth{0.150000\du}
\pgfsetdash{}{0pt}
\pgfsetdash{}{0pt}
\pgfsetbuttcap
{
\definecolor{dialinecolor}{rgb}{0.000000, 0.000000, 0.000000}
\pgfsetfillcolor{dialinecolor}
\pgfsetarrowsstart{to}
\pgfsetarrowsend{to}
\definecolor{dialinecolor}{rgb}{0.000000, 0.000000, 0.000000}
\pgfsetstrokecolor{dialinecolor}
\draw (37.000000\du,21.000000\du)--(52.000000\du,21.000000\du);
}
\definecolor{dialinecolor}{rgb}{0.000000, 0.000000, 0.000000}
\pgfsetstrokecolor{dialinecolor}
\node[anchor=west] at (44.000000\du,20.000000\du){\LARGE$C_d$};
\end{tikzpicture}

%% file: texFiles/numericalSimulations.tex
\section{Simulation Results}
\label{sec:simulationResults}

\subsection{Downlink and Channel Estimation Performance}
We compare the DL throughput and MSE performance of systems that employ the LS-based channel estimates obtained from RP to the performance of the massive MIMO systems that obtain channel estimates from SP and staggered pilots. 

Unless otherwise specified, the simulation parameters are as follows. The users are uniformly distributed in hexagonal cells and are at a distance of at least $ 100 $m from the BS. The inter-BS separation is $ 2 $km.  The channel estimation methods are tested with $ L = 91 $ cells and $ K = 5 $ users per cell. Both the SP and staggered pilots have reuse factors $ \pilotReuseRatioSp = 7 $. The path loss coefficient is assumed to be $ 3 $. The number of symbols in the UL and DL, i.e., $ \ulDuration $ and $ \dlDuration $, respectively, are both chosen as $ 35 $ symbols. The values of $ \rhoD{}{} $ and $ \rhoP{}{} $ are computed from \eqref{eqn:rhoDUlOptimal} and \eqref{eqn:rhoPUlOptimal}, respectively, and are used for both SP and staggered pilots. The UL transmit power $ \ulTotPower{\sIdxOne}{\sIdxTwo} $ is chosen based on the statistical channel-inversion power-control scheme \cite{bjornson2015massive}, i.e., $ \ulTotPower{\sIdxOne}{\sIdxTwo} = \powerBackoff/\beta_{\sIdxOne\sIdxOne\sIdxTwo} $ where $ \powerBackoff $ is a design parameter.
The signal-to-noise ratio (SNR) in the UL and DL, i.e., $ \powerBackoff/\sigma^2 $ and $ \dlSnr $, respectively, is set to $ 10 $dB. The plots are generated by averaging over $ 10^4 $ realizations of user locations across the cell. For the sake of simplicity, the effects of shadowing are not taken into account in this paper, but the conclusions are valid provided the users associate themselves with the strongest BS. 

\begin{figure}[t!]
	\centering
		\resizebox{0.55\columnwidth}{!}{\input{figs/fig2.tex}}
		\caption{Cumulative distribution of DL throughput for $ M = 100 $ antennas. The black line indicates rates with probability $ \geq 0.95 $. The DL rates for SP and staggered pilots that have a probability $ \leq 0.05 $ are significantly higher than those of RP with $ \pilotReuseRatioTp = 1 $.}
		\label{fig:cdfDlSinr}		
\end{figure}
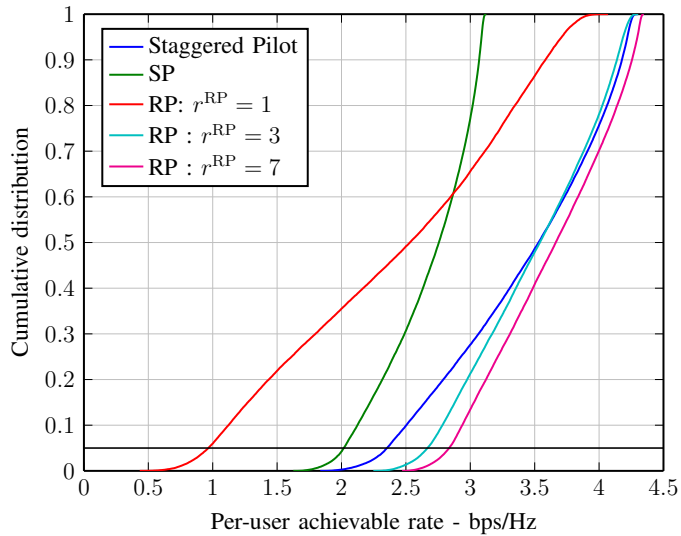

\begin{figure}[t!]
	\centering
	\resizebox{0.55\columnwidth}{!}{\input{figs/fig4.tex}}%
	\caption{DL Sum Rate vs $ M $ with MF precoder. SP and Staggered pilots offer an asymptotic DL throughput equivalent to that of RP with $ \pilotReuseRatioTp=7 $, even though the UL overhead is as much as that of RP with $ \pilotReuseRatioTp = 1 $.}%
	\label{fig:dlRate}%
	\vspace{0.5cm}
	\resizebox{0.55\columnwidth}{!}{\input{figs/fig6.tex}}%
	\caption{DL Sum Rate vs $ M $ with ZF precoder.}%
	\label{fig:dlRateZf}
\end{figure}

\begin{figure}[t!]
	\centering
		\resizebox{0.55\columnwidth}{!}{\input{figs/fig5.tex}}
		\caption{NMSE vs $ M $. Similar to Fig. \ref{fig:dlRate}, the asymptotic MSEs of SP and staggered pilots are equivalent to that of RP with $ \pilotReuseRatioTp = 7 $. In addition, the component of interference from users that transmit a pilot orthogonal to that of the reference user reduces asymptotically to zero.}
		\label{fig:mseVsM}	
\end{figure}
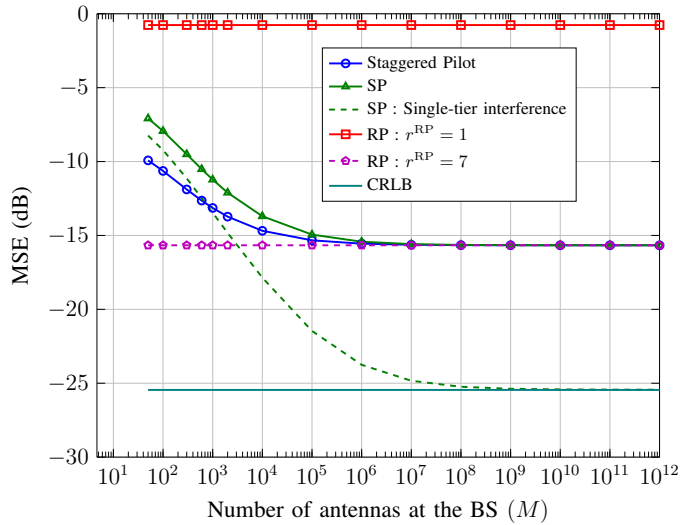

In Fig.~\ref{fig:cdfDlSinr}, the cumulative distribution of the DL rate of an arbitrary user in the reference BS is plotted for SP, RP, and staggered pilots for $ M = 100 $ antennas. The values of pilot reuse ratio for RP are $ \pilotReuseRatioTp = 1 $, $ \pilotReuseRatioTp = 3 $, and $ \pilotReuseRatioTp = 7 $ in the plot. It can be observed that rates obtained from the channel estimate based on SP and staggered pilots are significantly higher than those obtained from RP with $ \pilotReuseRatioTp = 1 $. Furthermore, it has to be noted that no additional UL overhead is required by staggered pilots and SP to achieve this DL throughput. However, while RP with $ \pilotReuseRatioTp = 3 $ offers comparable performance to staggered pilots with $ \pilotReuseRatioSp = 7 $, it has to be noted that, in addition to the increased UL overhead, pilot reuse with RP comes with the additional requirement that all users transmit their UL pilots simultaneously. This requirement will result in pilot reuse capable of being implemented across only a few cells in the network, and therefore, the remaining cells can benefit from using either SP or staggered pilots.

In Fig.~\ref{fig:dlRate}, the DL sum rate of staggered, RP, and SP is plotted against the number of antennas $ M $ when the channel estimates are used in an MF precoder. The DL rate ceiling of SP and staggered pilots is significantly higher than that of RP with $ \pilotReuseRatioTp=1 $. Moreover, the DL rate of staggered pilots is close to that of RP with $ \pilotReuseRatioTp=7 $ and SP achieves this rate asymptotically. The key difference between SP and staggered pilots is that in the former, the strongest interference seen by a particular user in the reference cell is due to the remaining users of that cell, which results from transmitting data alongside pilots. However, in the latter case, this component of interference is absent. Since the strongest component of the interference has been removed in the case of staggered pilots, they are capable of achieving a rate close to that of RP with $ \pilotReuseRatioTp = 7$. 

In Fig.~\ref{fig:dlRateZf}, the DL sum rates are plotted when the channel estimates are used in a ZF precoder which is given as
\begin{align}
\txPrecoder_{\vIdxOne\vIdxTwo} = \widehat{\mbf{H}}_{\vIdxOne}\lrc{\widehat{\mbf{H}}_{\vIdxOne}^H\widehat{\mbf{H}}_{\vIdxOne}}^{-1} \mbf{e}_{\vIdxTwo}
\end{align}
where $ \widehat{\mbf{H}}_{\vIdxOne} = \lrs{\widehat{\mbf{h}}_{\vIdxOne\vIdxOne0},\ldots,\widehat{\mbf{h}}_{\vIdxOne\vIdxOne,K-1}} $ and $ \mbf{e}_{\vIdxTwo} $ is the $ \vIdxTwo\rth $ column of $ \eye_{K} $. The parameter $ \txPrecoderNormalization_{\sIdxOne\sIdxTwo} $ is chosen to constrain the instantaneous transmit power to $ 1 $. The sum rates are obtained numerically with $ L = 7 $ and $ K = 5 $. It can be observed that the behavior of SP and RP is similar to that in Fig.~\ref{fig:dlRate}, whereas the gap between staggered pilots and SP/RP is significantly higher than that in Fig.~\ref{fig:dlRate}.

In Fig.~\ref{fig:mseVsM}, the MSE of the channel estimate is plotted against $ M $. Similar to the behavior in Fig.~\ref{fig:dlRate}, the MSE of the channel estimate obtained from SP and staggered pilots asymptotically approaches the MSE of the estimate from RP with $ \pilotReuseRatioTp = 7 $. In addition, since we have assumed that the interference from second and subsequent tiers of cells are negligible when deriving the CRLB, the interference from these cells results in a gap between the MSE of SP and the CRLB. In the absence of this component of the interference, it can be seen that the MSE of the channel estimate attains the CRLB asymptotically as $ M\rightarrow\infty $.

\begin{figure}[t!]
	\centering
		\resizebox{0.55\columnwidth}{!}{\input{figs/fig12.tex}}
		\caption{Sum rate in the UL over users in the first tier of cells vs. user radius. The algorithm assigns users RP and SP in scenarios with low and high interference, respectively.}
		\label{fig:hybridUlSumRateVsRadius}
	\vspace{1cm}
		\resizebox{0.55\columnwidth}{!}{\input{figs/fig13.tex}}
		\caption{Sum rate in the DL over users in the first tier of cells vs. user radius. The non-smooth nature of the DL rate of the hybrid system in the range of user radius $ [0.6,0.8] $ is due to the suboptimal nature of the greedy algorithm.}
		\label{fig:hybridDlSumRateVsRadius}
\end{figure}

\subsection{Hybrid System}
\label{sec:simulationResultsHybrid}
The hybrid system is simulated with $ L = 19 $ hexagonal cells, i.e., a central cell with two tiers of interfering cells. Each cell has $ K=5 $ users and the values of $ \ulDuration $ and $ \dlDuration $ are both chosen as $ 40 $ symbols. Although $ L $ is set to $ 19 $, the partitioning of users and the computation of the performance metrics is performed over $ 7 $ cells which consist of the central and the first tier of cells. The weights $ \hybridUlCostFactor $ and $ \hybridDlCostFactor $ are both set to $ 0.5 $. The value of $ \powerBackoff $ for the users in $ \spUserSet $ is set to $ 10 $ and $ \ulTotPower{}{} $ for the users in $ \cpUserSet $ is set to $ 1 $.

\renewcommand{\arraystretch}{1.5}
\begin{table}[t]
	\centering
	\caption{UL and DL performance of RP, SP, and hybrid systems}
	\label{table:hybridSystem}
	\newcolumntype{C}[2]{>{\hsize=#1\hsize\columncolor{#2}\centering\arraybackslash}X}%
	\begin{tabularx}{\linewidth}{p{4cm}XXXp{1.5cm}p{1.5cm}}
		\hline
		& UL Sum Rate (bps/Hz) & DL Sum Rate(bps/Hz) & Total Rate(bps/Hz)\\
		\hline
		Hybrid System & $ 48.12 $& $ 84.94 $& $ 133.07 $\\
		RPs $ (\pilotReuseRatioTp = 1) $ & $ 50.42 $ & $ 65.85 $ & $ 116.27 $ \\
		SPs & $ 35.30 $ &  $ 75.02 $ & $ 110.32 $ \\
		\hline
		
	\end{tabularx}
\end{table}

For obtaining Figs.~\ref{fig:hybridUlSumRateVsRadius} and \ref{fig:hybridDlSumRateVsRadius}, the users are assumed to be distributed uniformly on a circle around the BS. Then, the sum rates in the UL and DL are plotted in the figures against the radius of the circle around the BS. As can be observed in Figs. \ref{fig:hybridUlSumRateVsRadius} and \ref{fig:hybridDlSumRateVsRadius}, when the user radius is smaller than $ 0.6 $, RP are superior both in the UL and DL. However, SP are superior both in the UL and DL, when the user radius is larger than $ 0.8 $. Therefore, for user radius in the ranges $ [0,0.6] $ the greedy algorithm chooses RP and it chooses SP in the range $ [0.8,1] $. However, in the range $ [0.6,0.8] $, RP offer a better performance in the UL but a poorer performance in the DL, with respect to SP. Therefore, the choices of $ \cpUserSet $ and $ \spUserSet $ are determined by $ \hybridUlCostFactor $ and $ \hybridDlCostFactor $. Since for these simulations $ \hybridUlCostFactor $ and $ \hybridDlCostFactor $ are both chosen as $ 0.5 $, the greedy algorithm attempts to strike a balance between the UL and DL throughputs and offers a total throughput that is in between that of the systems that employ only RP or SP. In addition, since the algorithm is greedy, the variation of this throughput with respect to the user radius is non-smooth in nature, as can be seen in the figures.

Table \ref{table:hybridSystem} details the UL and DL performance of a system with users transmitting RP, SP, and a hybrid of both, when the users are uniformly distributed across the cells. 
The hybrid system offers roughly $ 14.44 \% $ higher total rate than the system that employs only RP. Moreover, both SP and the hybrid system offer a significantly higher throughput in the DL, albeit at the cost of a lower UL throughput than when compared with RP. However, the hybrid system enables controlling the trade-off between the UL and DL throughputs using the weights $ \hybridUlCostFactor $ and $ \hybridDlCostFactor $.

It has to be noted that there is an important difference between the results in Section \ref{sec:simulationResultsHybrid} and those in \cite{upadhya2016superimposed,upadhya2016hybrid}. In the latter, the computed rates are approximate for finite $ M $, since the correlation between the signal and interference components have been ignored and approximated to be zero. However, using the approach in Appendix \ref{appdx:dlSinr}, the signal and interference terms are uncorrelated and both the UL and DL rates shown in Figs.~\ref{fig:hybridUlSumRateVsRadius} and \ref{fig:hybridDlSumRateVsRadius}, and Table \ref{table:hybridSystem} are lower bounds on the achievable rates.

%% file: figs/fig2.tex
%
%
\definecolor{mycolor1}{rgb}{0.00000,0.75000,0.75000}%

\begin{tikzpicture}

\pgfplotsset{every axis legend/.append style={legend pos = north west},legend cell align = {left}}
\pgfplotsset{%
	tick label style={font=\normalsize},
	title style={font=\normalsize,align=center},
	label style={font=\normalsize,align=center},
	legend style={font=\normalsize}
}

\begin{axis}[%
width=4.0183in,
height=3.169in,
at={(0.809in,0.513in)},
ytick distance=0.1,
scale only axis,
separate axis lines,
every outer x axis line/.append style={black},
every x tick label/.append style={font=\color{black}},
every x tick/.append style={black},
xmin=0,
xmax=4.5,
xlabel=Per-user achievable rate - bps/Hz,
grid = major,
every outer y axis line/.append style={black},
every y tick label/.append style={font=\color{black}},
every y tick/.append style={black},
ymin=0,
ymax=1,
ylabel={Cumulative distribution},
axis background/.style={fill=white},
legend style={legend cell align=left, align=left, fill=white},
line width = 1pt
]

\addlegendentry{Staggered Pilot}

\addplot [ color = {blue}, 
mark  = {none},
mark options = {solid},
style = {solid},
line width = 1pt] table {figs/datFiles/fig2-line1.dat};

\addlegendentry{SP}

\addplot [ color = {black!50!green}, 
mark  = {none},
mark options = {solid},
style = {solid},
line width = 1pt] table {figs/datFiles/fig2-line2.dat};

\addlegendentry{RP: $\pilotReuseRatioTp = 1$}

\addplot [ color = {red}, 
mark  = {none},
mark options = {solid},
style = {solid},
line width = 1pt] table {figs/datFiles/fig2-line3.dat};

\addlegendentry{RP : $\pilotReuseRatioTp = 3$}

\addplot [ color = {mycolor1}, 
mark  = {none},
mark options = {solid},
style = {solid},
line width = 1pt] table {figs/datFiles/fig2-line4.dat};

\addlegendentry{RP : $\pilotReuseRatioTp = 7$}

\addplot [ color = {magenta}, 
mark  = {none},
mark options = {solid},
style = {solid},
line width = 1pt] table {figs/datFiles/fig2-line5.dat};
\draw [thick,black] (axis cs:-6,0.05) -- (axis cs:15,0.05);
\end{axis}
\end{tikzpicture}%

%% file: figs/fig4.tex
%
%
\definecolor{mycolor1}{rgb}{0.00000,0.75000,0.75000}%
\definecolor{mycolor2}{rgb}{0.75000,0.00000,0.75000}%
\begin{tikzpicture}
\pgfplotsset{every axis legend/.append style={at={(0.55,0.16)},anchor=west},legend cell align = {left}}
\pgfplotsset{%
	tick label style={font=\normalsize},
	title style={font=\normalsize,align=center},
	label style={font=\normalsize,align=center},
	legend style={font=\footnotesize}
}


\begin{semilogxaxis}[%
width=4.0183in,
height=3.169in,
at={(0.809in,0.513in)},
scale only axis,
separate axis lines,
every outer x axis line/.append style={black},
every x tick label/.append style={font=\color{black}},
every x tick/.append style={black},
grid = major,
xmin=1e0,
xmax=1e12,
xlabel={Number of antennas at the BS $(M)$},
every outer y axis line/.append style={black},
every y tick label/.append style={font=\color{black}},
every y tick/.append style={black},
ymin=0,
ymax=80,
ylabel={DL Sum Rate - bps/Hz},
axis background/.style={fill=white},
legend style={legend cell align=left, align=left, draw=black},
line width = 1pt
]
\addplot [color=blue,mark=o]
  table[row sep=crcr]{%
50	12.9517789398896\\
100	17.1508866036019\\
300	24.1721240700638\\
600	28.6353653014712\\
1000	31.8878544336284\\
2000	36.2170232542762\\
10000	45.7551708152754\\
100000	57.6953906811337\\
1000000	66.6866093034521\\
10000000	72.0060578681633\\
100000000	74.4514324621896\\
1000000000	75.3818476128674\\
10000000000	75.667742732645\\
100000000000	75.7372099421361\\
1000000000000	75.7542046443042\\
};
\addlegendentry{Staggered Pilot}

\addplot [color=black!50!green,mark=triangle]
  table[row sep=crcr]{%
50	10.3495857547318\\
100	13.3591530854829\\
300	18.2097738172877\\
600	21.2148253088688\\
1000	23.3785306738653\\
2000	26.2384763543748\\
10000	32.5702940831323\\
100000	41.0760784664746\\
1000000	49.0479597397485\\
10000000	56.2302471818689\\
100000000	62.1492256923207\\
1000000000	66.5813430332376\\
10000000000	69.7059273268291\\
100000000000	71.8480254045106\\
1000000000000	73.2938452764392\\
};
\addlegendentry{SP}

\addplot [color=red,mark=square]
  table[row sep=crcr]{%
50	9.26243033831638\\
100	12.1359731700606\\
300	16.3608028755001\\
600	18.6179798700526\\
1000	20.0549582979764\\
2000	21.7130300602127\\
10000	24.4748307144675\\
100000	26.6068114312462\\
1000000	27.5426143586943\\
10000000	27.8851199594241\\
100000000	27.9693281166109\\
1000000000	27.9813268686485\\
10000000000	27.9826032244282\\
100000000000	27.9827317346119\\
1000000000000	27.9827445946179\\
};
\addlegendentry{RP : $ \pilotReuseRatioTp = 1 $}

\addplot [color=mycolor1,mark=diamond]
  table[row sep=crcr]{%
50	13.1779159900088\\
100	17.4852666167284\\
300	24.720459775137\\
600	29.2800605793126\\
1000	32.5282183699349\\
2000	36.6630506974333\\
10000	44.4443343210701\\
100000	50.9062641663861\\
1000000	53.8390710857212\\
10000000	55.0889373712228\\
100000000	55.5613618175468\\
1000000000	55.6910413702077\\
10000000000	55.7119090822056\\
100000000000	55.7142155386055\\
1000000000000	55.7144489154316\\
};
\addlegendentry{RP : $ \pilotReuseRatioTp = 3 $}

\addplot [color=mycolor2,mark=pentagon,style=dashed,mark options=solid,line width = 2pt]
  table[row sep=crcr]{%
50	13.7286789468931\\
100	18.1351292575165\\
300	25.6196815775029\\
600	30.4866255736054\\
1000	34.0972544291272\\
2000	38.9911775749436\\
10000	50.041227474547\\
100000	63.0653857196475\\
1000000	70.4043555544518\\
10000000	73.6859411337753\\
100000000	75.0625580203158\\
1000000000	75.5851167408305\\
10000000000	75.7331295545787\\
100000000000	75.757991051355\\
1000000000000	75.7607853828413\\
};
\addlegendentry{RP : $ \pilotReuseRatioTp = 7	 $}

\end{semilogxaxis}
\end{tikzpicture}%

%% file: figs/fig6.tex
\begin{tikzpicture}

\pgfplotsset{every axis legend/.append style={legend pos = north west},legend cell align = {left}}
        
\pgfplotsset{%
tick label style={font=\normalsize},
title style={font=\normalsize,align=center},
label style={font=\normalsize,align=center},
legend style={font=\footnotesize}
            }

\begin{axis}[%
width=4.0183in,
height=3.169in,
at={(0.809in,0.513in)},
xlabel = {$M$},
ylabel = {DL Sum Rate - bps/Hz},
cycle list name = {color},
grid = major,
xmin = 150,
xmax = 600,
ymin = 10,
ymax = 40,
scale only axis,
scale = 1]

\addlegendentry{SP - ZF}

\addplot [ color = {black!50!green}, 
           mark  = {triangle},
           mark options = {solid},
           style = {solid},
           line width = 1pt] table {figs/datFiles/fig6-line1.dat};

\addlegendentry{RP : $\pilotReuseRatioTp$ = 1 - ZF}

\addplot [ color = {red}, 
           mark  = {square},
           mark options = {solid},
           style = {solid},
           line width = 1pt] table {figs/datFiles/fig6-line2.dat};

\addlegendentry{Staggered Pilot - ZF}

\addplot [ color = {blue}, 
           mark  = {o},
           mark options = {solid},
           style = {solid},
           line width = 1pt] table {figs/datFiles/fig6-line3.dat};

\addlegendentry{SP - MF}

\addplot [ color = {black!50!green}, 
mark  = {triangle},
mark options = {solid},
style = {dashed},
line width = 1pt] table {figs/datFiles/fig6-line4.dat};

\addlegendentry{RP : $\pilotReuseRatioTp$ = 1 - MF}

\addplot [ color = {red}, 
mark  = {square},
mark options = {solid},
style = {dashed},
line width = 1pt] table {figs/datFiles/fig6-line5.dat};

\addlegendentry{Staggered Pilot - MF}

\addplot [ color = {blue}, 
mark  = {o},
mark options = {solid},
style = {dashed},
line width = 1pt] table {figs/datFiles/fig6-line6.dat};

\end{axis}
\end{tikzpicture}

%% file: figs/fig5.tex
%
%
\definecolor{mycolor1}{rgb}{0.00000,0.75000,0.75000}%
\definecolor{mycolor2}{rgb}{0.75000,0.00000,0.75000}%
\definecolor{mycolor3}{rgb}{0.75000,0.75000,0.00000}%
\begin{tikzpicture}

\pgfplotsset{every axis legend/.append style={at={(0.4,0.75)},anchor=west},legend cell align = {left}}
\pgfplotsset{%
	tick label style={font=\normalsize},
	title style={font=\normalsize,align=center},
	label style={font=\normalsize,align=center},
	legend style={font=\footnotesize}
}
\begin{semilogxaxis}[%
width=4.0183in,
height=3.169in,
at={(0.809in,0.513in)},
scale only axis,
separate axis lines,
every outer x axis line/.append style={black},
every x tick label/.append style={font=\color{black}},
every x tick/.append style={black},
grid=major,
xmin=0,
xmax=1000000000000,
xlabel={Number of antennas at the BS $(M)$},
every outer y axis line/.append style={black},
every y tick label/.append style={font=\color{black}},
every y tick/.append style={black},
ymin=-30,
ymax=0,
ylabel={MSE (dB)},
axis background/.style={fill=white},
legend style={legend cell align=left, align=left, draw=black}
]
\addplot [color=blue, line width = 1pt,mark=o]
  table[row sep=crcr]{%
50	-9.92141272479294\\
100	-10.6348562657331\\
300	-11.8873116942071\\
600	-12.6394603816122\\
1000	-13.1395295424175\\
2000	-13.7269761452311\\
10000	-14.6849388372924\\
100000	-15.3306258823947\\
1000000	-15.5570622246765\\
10000000	-15.6312130969103\\
100000000	-15.6549281435199\\
1000000000	-15.6624545472162\\
10000000000	-15.6648373226491\\
100000000000	-15.6655910945707\\
1000000000000	-15.6658294854114\\
};
\addlegendentry{Staggered Pilot}

\addplot [color=black!50!green, line width = 1pt,mark=triangle]
  table[row sep=crcr]{%
50	-7.08350183247516\\
100	-7.93418357380895\\
300	-9.50573810313001\\
600	-10.5152785118162\\
1000	-11.2237687440009\\
2000	-12.1052709316433\\
10000	-13.7013978267657\\
100000	-14.9428202948681\\
1000000	-15.4239590279265\\
10000000	-15.5879499984725\\
100000000	-15.6411254310647\\
1000000000	-15.6580774258023\\
10000000000	-15.6634519185121\\
100000000000	-15.6651528674831\\
1000000000000	-15.6656908934505\\
};
\addlegendentry{SP}

\addplot [color=black!50!green, line width = 1pt,style=dashed]
  table[row sep=crcr]{%
50	-8.24226745311014\\
100	-9.2253503907359\\
300	-11.1395918330259\\
600	-12.4678774955083\\
1000	-13.4685274628507\\
2000	-14.8271184801329\\
10000	-17.8544063113714\\
100000	-21.4611287764697\\
1000000	-23.7543977740566\\
10000000	-24.8335709511512\\
100000000	-25.239367265215\\
1000000000	-25.3760050571647\\
10000000000	-25.4201239709888\\
100000000000	-25.4341693905363\\
1000000000000	-25.4386204122171\\
};
\addlegendentry{SP : Single-tier interference}

\addplot [color=red, line width = 1pt,mark=square]
  table[row sep=crcr]{%
50	-0.76368002344525\\
100	-0.76368002344525\\
300	-0.76368002344525\\
600	-0.76368002344525\\
1000	-0.76368002344525\\
2000	-0.76368002344525\\
10000	-0.76368002344525\\
100000	-0.76368002344525\\
1000000	-0.76368002344525\\
10000000	-0.76368002344525\\
100000000	-0.76368002344525\\
1000000000	-0.76368002344525\\
10000000000	-0.76368002344525\\
100000000000	-0.76368002344525\\
1000000000000	-0.76368002344525\\
};
\addlegendentry{RP : $ \pilotReuseRatioTp = 1 $}


\addplot [color=mycolor2,line width = 1pt,mark=pentagon,style=dashed,mark options=solid]
  table[row sep=crcr]{%
50	-15.6659397397114\\
100	-15.6659397397114\\
300	-15.6659397397114\\
600	-15.6659397397114\\
1000	-15.6659397397114\\
2000	-15.6659397397114\\
10000	-15.6659397397114\\
100000	-15.6659397397114\\
1000000	-15.6659397397114\\
10000000	-15.6659397397114\\
100000000	-15.6659397397114\\
1000000000	-15.6659397397114\\
10000000000	-15.6659397397114\\
100000000000	-15.6659397397114\\
1000000000000	-15.6659397397114\\
};
\addlegendentry{RP : $ \pilotReuseRatioTp= 7 $}

\addplot [color=teal,line width = 1pt]
  table[row sep=crcr]{%
50	-25.4530711646582\\
100	-25.4530711646582\\
300	-25.4530711646582\\
600	-25.4530711646582\\
1000	-25.4530711646582\\
2000	-25.4530711646582\\
10000	-25.4530711646582\\
100000	-25.4530711646582\\
1000000	-25.4530711646582\\
10000000	-25.4530711646582\\
100000000	-25.4530711646582\\
1000000000	-25.4530711646582\\
10000000000	-25.4530711646582\\
100000000000	-25.4530711646582\\
1000000000000	-25.4530711646582\\
};
\addlegendentry{CRLB}

\end{semilogxaxis}
\end{tikzpicture}%

%% file: figs/fig12.tex
\begin{tikzpicture}[scale = 1]

\pgfplotsset{every axis legend/.append style={legend pos = north east},legend cell align = {left}}
        
\pgfplotsset{%
tick label style={font=\footnotesize},
title style={font=\normalsize,align=center},
label style={font=\footnotesize,align=center},
legend style={font=\footnotesize}
            }

\begin{axis}[%
xlabel = {Radius},
ylabel = {UL Sum Rate - bps/Hz},
enlargelimits = true,
cycle list name = {color},
grid = major,
xmin = 0.2,
xmax = 0.9,
ymin = 11,
ymax = 125,
scale = 1,
xtick distance=0.1,
ytick distance=20,
]

\addlegendentry{RP}

\addplot [ color = {cyan}, 
           mark  = {square},
           mark options = {solid},
           style = {solid},
           line width = 1pt] table {figs/datFiles/fig12-line2.dat};

\addlegendentry{SP}

\addplot [ color = {red}, 
           mark  = {o},
           mark options = {solid},
           style = {solid},
           line width = 1pt] table {figs/datFiles/fig12-line3.dat};

\addlegendentry{Hybrid System}

\addplot [ color = {violet}, 
mark  = {pentagon},
mark options = {solid},
style = {solid},
line width = 1pt] table {figs/datFiles/fig12-line1.dat};

\end{axis}
\end{tikzpicture}

%% file: figs/fig13.tex
\begin{tikzpicture}[scale = 1]

\pgfplotsset{every axis legend/.append style={legend pos = north east},legend cell align = {left}}
        
\pgfplotsset{%
tick label style={font=\footnotesize},
title style={font=\normalsize,align=center},
label style={font=\footnotesize,align=center},
legend style={font=\footnotesize}
            }

\begin{axis}[%
xlabel = {Radius},
ylabel = {DL Sum Rate - bps/Hz},
enlargelimits = true,
cycle list name = {color},
grid = major,
scale = 1,
xmin = 0.2,
xmax = 0.9,
ymin = 11,
ymax = 125,
xtick distance=0.1]

\addlegendentry{RP}

\addplot [ color = {cyan}, 
           mark  = {square},
           mark options = {solid},
           style = {solid},
           line width = 1pt] table {figs/datFiles/fig13-line2.dat};

\addlegendentry{SP}

\addplot [ color = {red}, 
           mark  = {o},
           mark options = {solid},
           style = {solid},
           line width = 1pt] table {figs/datFiles/fig13-line3.dat};

\addlegendentry{Hybrid System}

\addplot [ color = {violet}, 
mark  = {pentagon},
mark options = {solid},
style = {solid},
line width = 1pt] table {figs/datFiles/fig13-line1.dat};
        
\end{axis}
\end{tikzpicture}

%% file: texFiles/conclusion.tex
\section{Conclusion}
\label{sec:conclusion}
We have shown that SPs offer a significantly better asymptotic MSE and DL performance than RPs. This improvement is attributed to utilizing the array gain of the antenna for reducing the fraction of UL power allocated to data in favor of allocating a larger fraction of power for pilot transmission. We have also shown that staggered pilots are a particular case of SPs and therefore, offer a DL performance similar to that of SPs, while offering the same UL spectral and energy efficiency as RPs. Furthermore, we have shown that higher asymptotic DL throughput offered by SPs and staggered pilots are at the same or lower UL transmission overhead than RPs. We have also extended the hybrid system to partition the users into two disjoint sets of users that use RPs and SPs by minimizing both the UL and DL interference. We show, by means of simulation, that the hybrid system offers a higher throughput than when only RPs or SPs are employed.


%% file: texFiles/appendixDlSinrCalcFeedback.tex
\appendices
\section{}
\label{appdx:dlSinr}
\subsection*{Downlink SINR for Channel Estimate Based on SP Pilots}

For MF precoding, $ \txPrecoder_{\sIdxOne\sIdxTwo} = \widehat{\mbf{h}}_{\sIdxOne\sIdxOne\sIdxTwo} $. Then,
\begin{align}
	\expectation
	\lrf{			
		\mbf{h}_{\vIdxOne\vIdxOne\vIdxTwo}^H
		\txPrecoder_{\vIdxOne\vIdxTwo}
	}	
	&=
	M\beta_{\vIdxOne\vIdxOne\vIdxTwo}
	\label{eqn:interfTerm1}
	\\
	\expectation
	\lrf{
		|
		\mbf{h}_{\sIdxOne\vIdxOne\vIdxTwo}^H
		\txPrecoder_{\sIdxOne\sIdxTwo}	
		|^2}
	&=
	M^2
	\beta_{\sIdxOne\vIdxOne\vIdxTwo}^2
	\frac{\ulTotPower{\vIdxOne}{\vIdxTwo}}{\ulTotPower{\sIdxOne}{\vIdxTwo}}
	\indicator{\sIdxTwo=\vIdxTwo,\vIdxOne\in\cellSharingPilotSetSp{\sIdxOne}}
	+
	M
	\sum\limits_{\sIdxThree\in\cellSharingPilotSetSp{\sIdxOne}}
	\frac{\ulTotPower{\sIdxThree}{\sIdxTwo}}{\ulTotPower{\sIdxOne}{\sIdxTwo}}
	\beta_{\sIdxOne\vIdxOne\vIdxTwo}
	\beta_{\sIdxOne\sIdxThree\sIdxTwo}
	\nonumber
	\\
	&
	+
	M
	\sum\limits_{\sIdxThree=0}^{L-1}
	\sum\limits_{\sIdxFour=0}^{K-1}
	\frac{\rhoD{}{}^2}{\ulDuration\rhoP{}{}^2}
	\frac{\ulTotPower{\sIdxThree}{\sIdxFour}}{\ulTotPower{\sIdxOne}{\sIdxTwo}}
	\beta_{\sIdxOne\sIdxThree\sIdxFour}
	\beta_{\sIdxOne\vIdxOne\vIdxTwo}
	+
	\frac{M^2\rhoD{}{}^2}{\ulDuration\rhoP{}{}^2}
	\frac{\ulTotPower{\vIdxOne}{\vIdxTwo}}{\ulTotPower{\sIdxOne}{\sIdxTwo}}
	\beta_{\sIdxOne\vIdxOne\vIdxTwo}^2
	+
	\frac{M\sigma^2}{\ulDuration\rhoP{}{}^2\ulTotPower{\sIdxOne}{\sIdxTwo	}}
	\\
	\frac{1}{\txPrecoderNormalization_{\sIdxOne\sIdxTwo}}
	=
	\frac{\expectation
	\lrf{\|\txPrecoder_{\sIdxOne\sIdxTwo}\|^2}}{\dlTxPower_{\sIdxOne\sIdxTwo}}
	&=
	\frac{M}{\dlTxPower_{\sIdxOne\sIdxTwo}}
	\lrc{
	\sum\limits_{\sIdxThree\in\cellSharingPilotSetSp{\sIdxOne}}
	\frac{\ulTotPower{\sIdxThree}{\sIdxTwo}}{\ulTotPower{\sIdxOne}{\sIdxTwo}}
	\beta_{\sIdxOne\sIdxThree\sIdxTwo}
	+
	\sum\limits_{\sIdxThree=0}^{L-1}	
	\sum\limits_{\sIdxFour=0}^{K-1}	
	\frac{\rhoD{}{}^2}{\ulDuration\rhoP{}{}^2}
	\frac{\ulTotPower{\sIdxThree}{\sIdxFour}}{\ulTotPower{\sIdxOne}{\sIdxTwo}}
	\beta_{\sIdxOne\sIdxThree\sIdxFour}
	+
	\frac{\sigma^2}{\ulDuration\rhoP{}{}^2\ulTotPower{\sIdxOne}{\sIdxTwo}}
	}
	\label{eqn:interfTerm3}
\end{align}
Substituting \eqref{eqn:interfTerm1} to \eqref{eqn:interfTerm3} into \eqref{eqn:interferencePowerExpansion}, the SINR can be obtained as
\begin{align}
	\spDlSinr_{\vIdxOne\vIdxTwo}
	\!
	&=
	\!
	\txPrecoderNormalization_{\vIdxOne\vIdxTwo}\beta_{\vIdxOne\vIdxOne\vIdxTwo}^2
	\!\!
	\left(
		\sum\limits_{\cellSharingPilotSetSp{\vIdxOne}\ni\sIdxOne\neq\vIdxOne}
		\!
		\frac{\ulTotPower{\vIdxOne}{\vIdxTwo}}{\ulTotPower{\sIdxOne}{\vIdxTwo}}
		\txPrecoderNormalization_{\vIdxOne\vIdxTwo}
		\beta_{\sIdxOne\vIdxOne\vIdxTwo}^2
		+
		\frac{1}{M}
		\sum\limits_{\sIdxOne=0}^{L-1}
		\sum\limits_{\sIdxTwo=0}^{K-1}
		\frac{\txPrecoderNormalization_{\sIdxOne\sIdxTwo}\beta_{\sIdxOne\vIdxOne\vIdxTwo}}{\ulTotPower{\sIdxOne}{\sIdxTwo}}		
		\!\!
		\lrc{
			\sum\limits_{\sIdxThree\in\cellSharingPilotSetSp{\sIdxOne}}		
			\ulTotPower{\sIdxThree}{\sIdxTwo}
			\beta_{\sIdxOne\sIdxThree\sIdxTwo}
			+
			\frac{\sigma^2}{\ulDuration\rhoP{}{}^2}
			\!
		}
		\right.
		\nonumber
		\\
		&\left.
		+
		\frac{\rhoD{}{}^2}{\ulDuration\rhoP{}{}^2}
		\sum\limits_{\sIdxOne=0}^{L-1}
		\sum\limits_{\sIdxTwo=0}^{K-1}		
		\frac{\txPrecoderNormalization_{\sIdxOne\sIdxTwo}}{\ulTotPower{\sIdxOne}{\sIdxTwo}}
		\lrc{
			\ulTotPower{\vIdxOne}{\vIdxTwo}
			\beta_{\sIdxOne\vIdxOne\vIdxTwo}^2
			+
		\frac{1}{M}
		\sum\limits_{\sIdxThree=0}^{L-1}		
		\sum\limits_{\sIdxFour=0}^{K-1}	
		\ulTotPower{\sIdxThree}{\sIdxFour}			
		\beta_{\sIdxOne\sIdxThree\sIdxFour}
		\beta_{\sIdxOne\vIdxOne\vIdxTwo}
		}
		+
		\frac{1}{M^2\dlSnr}
	\right)^{-1}
\end{align}
This completes the derivation of \eqref{eqn:spDlSinrMf}.

%% file: texFiles/appendixRhoDRhoPCalc.tex
\section{}
\subsection*{Calculation of $ \rhoD{}{}^2_{\mathrm{opt}} $ and $ \rhoP{}{}^2_{\mathrm{opt}} $}
\label{appdx:optRhoDRhoP}

Equation \eqref{eqn:ulSystemModel} can be written as
\begin{align}
\widetilde{\mbf{Y}}_{\vIdxOne}
=
\sum\limits_{\sIdxOne=0}^{L-1}
\sum\limits_{\sIdxTwo=0}^{K-1}
\sqrt{\ulTotPower{\sIdxOne}{\sIdxTwo}}
\mbf{h}_{\vIdxOne\sIdxOne\sIdxTwo}
\lrc{
	\rhoP{}{}
	\spPilot{\sIdxOne}{\sIdxTwo}
	+
	\rhoD{}{}
	\widetilde{\mbf{x}}_{\sIdxOne\sIdxTwo}
}^T
+
\widetilde{\mbf{W}}_{\vIdxOne}
\label{eqn:systemModel}
\end{align}
where $ \widetilde{\mbf{x}}_{\sIdxOne\sIdxTwo} \sim \mathcal{CN}(\mbf{0},\eye) $. The receiver applies the following linear invertible transformation to the received observation 
\begin{align}
\mbf{Y}_{\vIdxOne}
\triangleq
\frac{1}{\sqrt{\ulDuration}}
\widetilde{\mbf{Y}}_{\vIdxOne}
\mbf{P}^*
&=	
\sum\limits_{\sIdxOne=0}^{L-1}
\sum\limits_{\sIdxTwo=0}^{K-1}
\sqrt{\ulTotPower{\sIdxOne}{\sIdxTwo}}
\mbf{h}_{\vIdxOne\sIdxOne\sIdxTwo}
\lrc{		
	\rhoP{}{}\sqrt{\ulDuration}
	\mbf{e}_{\sIdxOne\sIdxTwo}
	+
	\rhoD{}{}
	\mbf{x}_{\sIdxOne\sIdxTwo}		
}^T	
+
\mbf{W}_{\vIdxOne}		
\label{eqn:systemModelOrthogonalized}
\end{align}
where $ \mbf{e}_{\sIdxOne\sIdxTwo} = \mbf{P}^H\spPilot{\sIdxOne}{\sIdxTwo} /\ulDuration $ has ones in locations corresponding to the column index of $ \spPilot{\sIdxOne}{\sIdxTwo} $ in $ \mbf{P} $, $ \mbf{x}_{\sIdxOne\sIdxTwo} = \mbf{P}^H\widetilde{\mbf{x}}_{\sIdxOne\sIdxTwo}/\sqrt{\ulDuration} $, and $ \mbf{W}_{\vIdxOne} = \widetilde{\mbf{W}} \mbf{P}^H / \sqrt{\ulDuration} $. Note that the distributions of $ \mbf{x}_{\sIdxOne\sIdxTwo} $ and $ \mbf{W}_{\vIdxOne} $ are unchanged since $ \mbf{P} / \sqrt{\ulDuration} $ is a unitary matrix. Therefore, from the perspective of calculating the achievable rate, both \eqref{eqn:systemModel} and \eqref{eqn:systemModelOrthogonalized} are equivalent. We simplify notation by dropping the subscript $ \vIdxOne $ and replacing the tuple $ \lrc{\sIdxOne,\sIdxTwo} $ with a single index $ \sIdxTwo $ such that $ 0\leq\sIdxTwo\leq N-1 $, where $ N \triangleq LK-1 $. Defining $ \mbf{y}_{\symbolIdx} $ as the $ \symbolIdx\rth $ vector of received observations and $ x_{\userIdx\symbolIdx} $ as the $ \symbolIdx\rth $ element of $ \mbf{x}_{\userIdx} $, $ \mbf{y}_{\symbolIdx} $ for $ 0\leq\symbolIdx\leq\ulDuration-1 $ can be written as
\begin{align}
\mbf{y}_{\symbolIdx}
=
\sum\limits_{\sIdxTwo \in \pilotSharingSet_{\symbolIdx}}
\rhoP{}{}
\sqrt{\ulTotPower{\sIdxTwo}{}\ulDuration}
\mbf{h}_{\sIdxTwo}
+
\sum\limits_{\sIdxTwo=0}^{N-1}
\rhoD{}{}
\sqrt{\ulTotPower{\sIdxTwo}{}}
\mbf{h}_{\sIdxTwo}
x_{\sIdxTwo\symbolIdx}
+
\mbf{w}_{\symbolIdx}
\label{eqn:spEqvtUL}
\end{align}
where $ \pilotSharingSet_{\symbolIdx} = \lrf{\sIdxTwo \in \lrf{0,\ldots,LK-1} \;|\; \mbf{p}_{\sIdxTwo} = \mbf{p}_{\symbolIdx}} $ is the set of all users that transmit their pilot in symbol $ \symbolIdx $, and $ \pilotSharingSet_{\symbolIdx} = \varnothing $ for $ \symbolIdx \geq LK $.

Without loss of generality, let user $ 0 $ be the reference user, and let this user transmit its pilot in symbol $ 0 $ in \eqref{eqn:spEqvtUL}. The LS estimate of the channel of user $ 0 $ can then be written as
\begin{align}
\widehat{\mbf{h}}_{0} 
&= 
\frac{1}{\sqrt{\ulTotPower{0}{}}\sqrt{\ulDuration}\rhoP{}{}}
\mbf{y}_{0}
=
\sum\limits_{\sIdxTwo \in \pilotSharingSet_{0}}
\sqrt{\frac{\ulTotPower{\sIdxTwo}{}}{\ulTotPower{0}{}}}
\mbf{h}_{\sIdxTwo}
+
\sum\limits_{\sIdxTwo=0}^{N-1}
\frac{\rhoD{}{}}{\sqrt{\ulDuration}\rhoP{}{}}
\sqrt{\frac{\ulTotPower{\sIdxTwo}{}}{\ulTotPower{0}{}}}
\mbf{h}_{\sIdxTwo}
x_{\sIdxTwo 0}
+
\frac{\mbf{w}_{\symbolIdx}}{\sqrt{\ulDuration\ulTotPower{0}{}}\rhoP{}{}}\;.
\label{eqn:lsChannelEstimateOrthogonalized}
\end{align}
Then, the output of the MRC when $ \symbolIdx \geq 1 $ is
\begin{align}
\widehat{x}_{0\symbolIdx}
=
\widehat{\mbf{h}}_{0}^H
\mbf{y}_{\symbolIdx}
&=
\rhoD{}{}
\sqrt{\ulTotPower{0}{}}
\expectation
\lrf{
	\widehat{\mbf{h}}_{0}^H
	\mbf{h}_{0}
}
x_{0\symbolIdx}
+
\rhoD{}{}
\sqrt{\ulTotPower{0}{}}
\lrc{
	\widehat{\mbf{h}}_{0}^H
	\mbf{h}_{0}
	-
	\expectation
	\lrf{
		\widehat{\mbf{h}}_{0}^H
		\mbf{h}_{0}
	}
}
x_{0\symbolIdx}
+
\sum\limits_{\sIdxTwo \in \pilotSharingSet_{\symbolIdx}}
\rhoP{}{}
\sqrt{\ulTotPower{\sIdxTwo}{}\ulDuration}
\widehat{\mbf{h}}_{0}^H
\mbf{h}_{\sIdxTwo}
\nonumber
\\
&\qquad
+
\sum\limits_{\sIdxTwo=1}^{N-1}
\rhoD{}{}
\sqrt{\ulTotPower{\sIdxTwo}{}}
\widehat{\mbf{h}}_{0}^H
\mbf{h}_{\sIdxTwo}
x_{\sIdxTwo\symbolIdx}
+
\widehat{\mbf{h}}_{0}^H
\mbf{w}_{\symbolIdx}\;.
\label{eqn:dataEstimate}
\end{align}
Since $ x_{0\symbolIdx} $ when $ \symbolIdx \geq 1 $ is independent of $x_{\userIdx 0} $ for all $ 0\leq\userIdx\leq LK-1 $ , $ x_{0\symbolIdx} $ is independent of $ \widehat{\mbf{h}}_{0} $. As a result, in \eqref{eqn:dataEstimate}, the first term is uncorrelated with the remaining terms. Then, \eqref{eqn:dataEstimate} can be written as
$
\widehat{x}_{0\symbolIdx}
=
s_{\symbolIdx}
+
i_{\symbolIdx}
$
where
\begin{align}
s_{\symbolIdx}
&\triangleq
\rhoD{}{}
\sqrt{\ulTotPower{0}{}}
\expectation
\lrf{
	\widehat{\mbf{h}}_{0}^H
	\mbf{h}_{0}
}
x_{0\symbolIdx}
\\
i_{\symbolIdx}
&\triangleq
\rhoD{}{}
\sqrt{\ulTotPower{0}{}}
\lrc{
	\widehat{\mbf{h}}_{0}^H
	\mbf{h}_{0}
	-
	\expectation
	\lrf{
		\widehat{\mbf{h}}_{0}^H
		\mbf{h}_{0}
	}
}
x_{0\symbolIdx}
+
\sum\limits_{\sIdxTwo \in \pilotSharingSet_{\symbolIdx}}
\rhoP{}{}
\sqrt{\ulTotPower{\sIdxTwo}{}\ulDuration}
\widehat{\mbf{h}}_{0}^H
\mbf{h}_{\sIdxTwo}
+
\sum\limits_{\sIdxTwo=1}^{N-1}
\rhoD{}{}
\sqrt{\ulTotPower{\sIdxTwo}{}}
\widehat{\mbf{h}}_{0}^H
\mbf{h}_{\sIdxTwo}
x_{\sIdxTwo\symbolIdx}
+
\widehat{\mbf{h}}_{0}^H
\mbf{w}_{\symbolIdx}\;.
\end{align}
Defining $ \logCapacity{x} = \log_2\lrc{1+x} $, a lower bound on the ergodic capacity for user $ 0 $ in symbol $ \symbolIdx $ can be obtained as \cite{Hassibi2003HowMuch}
\begin{align}
R_{0\symbolIdx}
=
\logCapacity{\frac{\expectation\lrf{|s_{\symbolIdx}|^2}}{\expectation\lrf{|i_{\symbolIdx}|^2}}}
\label{eqn:capacitySingleSymbol} \;.
\end{align}
Since $ \signalPower\triangleq\expectation\lrf{|s_{\symbolIdx}|^2} $ is independent of $ \symbolIdx $ for $ \symbolIdx \geq 1 $ and since $ \log_2\lrc{1+1/x} $ is convex in $ x $, an achievable lower bound on the capacity can be obtained using Jensen's inequality as
\begin{align}
R_{0}
&=
\frac{1}{\ulDuration}
\!\!
\sum\limits_{\symbolIdx=0}^{\ulDuration-1}
\!\!
\logCapacity{\frac{\signalPower}{\expectation\lrf{|i_{\symbolIdx}|^2}}}
\!\!
\geq
\!
\frac{1}{\ulDuration}
\sum\limits_{\symbolIdx=1}^{\ulDuration-1}
\logCapacity{\frac{\signalPower}{\expectation\lrf{|i_{\symbolIdx}|^2}}}
\!
\geq
\frac{\ulDuration-1}{\ulDuration}
\logCapacity{\frac{\signalPower}{\frac{1}{\ulDuration-1}\sum\limits_{\symbolIdx=1}^{\ulDuration-1}\expectation\lrf{|i_{\symbolIdx}|^2}}}
\label{eqn:achRateFullFrame}
\end{align}
where, in the first inequality, the throughput in symbol $ 0 $ in which user $ 0 $ transmits both pilot and data is ignored. Jensen's inequality is applied in the second inequality to render the right-hand side independent of $ \symbolIdx $. Now,
\begin{align}
\interferencePower
&\triangleq
\frac{1}{\ulDuration-1}
\sum\limits_{\symbolIdx=1}^{\ulDuration-1}
\expectation\lrf{|i_{\symbolIdx}|^2}
=
\sum\limits_{\sIdxTwo=0}^{N-1}
\rhoD{}{}^2
\ulTotPower{\sIdxTwo}{}
\expectation
\lrf{
	|
	\widehat{\mbf{h}}_{0}^H
	\mbf{h}_{\sIdxTwo}
	|^2
}
-
\rhoD{}{}^2
\ulTotPower{0}{}
\left|		
\expectation
\lrf{
	\widehat{\mbf{h}}_{0}^H
	\mbf{h}_{0}
}
\right|^2
\nonumber
\\
&+
\frac{1}{\ulDuration-1}
\sum\limits_{\symbolIdx=1}^{\ulDuration-1}
\sum\limits_{\sIdxTwo \in \pilotSharingSet_{\symbolIdx}}
\ulTotPower{\sIdxTwo}{}
\rhoP{}{}^2
\ulDuration
\expectation
\lrf{
	|
	\widehat{\mbf{h}}_{0}^H
	\mbf{h}_{\sIdxTwo}
	|^2
}
+
\expectation
\lrf{
	|
	\widehat{\mbf{h}}_{0}^H
	\mbf{w}_{\symbolIdx}
	|^2
}
\nonumber
\\
&\leq
\frac{\ulDuration}{\ulDuration-1}
\sum\limits_{\sIdxTwo=0}^{N-1}
\ulTotPower{\sIdxTwo}{}
\expectation
\lrf{
	|
	\widehat{\mbf{h}}_{0}^H
	\mbf{h}_{\sIdxTwo}
	|^2
}
-
\rhoD{}{}^2
\ulTotPower{0}{}
\left|		
\expectation
\lrf{
	\widehat{\mbf{h}}_{0}^H
	\mbf{h}_{0}
}
\right|^2
+
\expectation
\lrf{
	|
	\widehat{\mbf{h}}_{0}^H
	\mbf{w}_{\symbolIdx}
	|^2
}
\label{eqn:interferencePower}
\end{align}
where, to obtain the inequality, we have used the property that $\bigcup\limits_{\symbolIdx=1}^{\ulDuration-1} \pilotSharingSet_{\symbolIdx} = \userSet\backslash \pilotSharingSet_{0} \subset \userSet $ with $ \userSet \triangleq  \lrf{0,\ldots,N-1} $ being the set of all users, and that $ \ulDuration / (\ulDuration-1) > 1 $. Using \eqref{eqn:lsChannelEstimateOrthogonalized}, $ \expectation
\lrf{
	|
	\widehat{\mbf{h}}_{0}^H
	\mbf{h}_{\sIdxTwo}
	|^2
}
$ in \eqref{eqn:interferencePower} can be obtained as
\begin{align}
	\expectation\lrf{|\widehat{\mbf{h}}_{0}^H\mbf{h}_{\sIdxTwo}|^2}	
	=
	\begin{cases}
		\tTerm_{\sIdxTwo} 
		& 
		\sIdxTwo \notin \pilotSharingSet_{0} \\
		M^2		
		\beta_{\sIdxTwo}^2
		\frac{\ulTotPower{\sIdxTwo}{}}{\ulTotPower{0}{}}
		+
		\tTerm_{\sIdxTwo}
		&
		\sIdxTwo \in \pilotSharingSet_{0}
	\end{cases}
\end{align}
where
\begin{align}
	\tTerm_{\sIdxTwo} 
	&\triangleq
	M^2
	\rhoD{}{}^2
	\lrc{
		\frac{c_{\sIdxTwo}}{\ulDuration\rhoP{}{}^2}		
		+
		\frac{1}{M\rhoD{}{}^2}
		d_{\sIdxTwo}
		+
		\frac{1}{M\rhoD{}{}^2\rhoP{}{}^2}
		e_{\sIdxTwo}
	}
	\label{eqn:tTermDefn}
	\\
	c_{\sIdxTwo}
	&\triangleq
		\frac{\ulTotPower{\sIdxTwo}{}}{\ulTotPower{0}{}}	
		\beta_{\sIdxTwo}^2
		+
		\frac{1}{M}
		\sum\limits_{\sIdxThree=0}^{N-1}	
		\frac{\ulTotPower{\sIdxThree}{}}{\ulTotPower{0}{}}
		\beta_{\sIdxThree}
		\beta_{\sIdxTwo}
	\;,\;
	d_{\sIdxTwo}
	\triangleq
	\sum\limits_{\sIdxThree \in \pilotSharingSet_{0}}
	\frac{\ulTotPower{\sIdxThree}{}}{\ulTotPower{0}{}}
	\beta_{\sIdxThree}
	\beta_{\sIdxTwo}	
	\;,\;
	e_{\sIdxTwo}
	\triangleq
	\frac{\beta_{\sIdxTwo}\sigma^2}{\ulTotPower{0}{}\ulDuration}	
\end{align}
Substituting \eqref{eqn:tTermDefn} into \eqref{eqn:interferencePower} and noting that $ \mbf{w}_{\symbolIdx} $ is independent of $ \widehat{\mbf{h}}_{0} $, we obtain
\begin{align}
	\interferencePower
	&\leq
	M^2
	\rhoD{}{}^2
	\lrs{
	\frac{\rhoD{}{}^2}{\rhoP{}{}^2}	
	\cValOne
	+
	\frac{\rhoD{}{}^2}{M\rhoP{}{}^2}	
	\cValTwo
	+
	\frac{
		\rhoP{}{}^2
	}{\rhoD{}{}^2}	
	\frac{\cValThree}{M}
	+	
	\frac{1}{\rhoD{}{}^2\rhoP{}{}^2}
	\frac{\cValFour}{M}
	+
	\frac{1}{\rhoD{}{}^2}
	\frac{\cValFive}{M}
	+
	\frac{1}{\rhoP{}{}^2}
	\frac{\cValSix}{M}
	+
	\cValSeven
	}
	\label{eqn:interfPowerUpperBoundAlphaSum}
\end{align}
where 
\begin{align}
\cValOne
&=
\sum\limits_{\sIdxTwo=0}^{N-1}	
\frac{\ulTotPower{\sIdxTwo}{}^2\beta_{\sIdxTwo}^2}{\ulTotPower{0}{}(\ulDuration-1)}
\;,\;
\cValTwo
=
\sum\limits_{\sIdxTwo=0}^{N-1}	
\sum\limits_{\sIdxThree=0}^{N-1}	
\frac{\ulTotPower{\sIdxTwo}{}\ulTotPower{\sIdxThree}{}\beta_{\sIdxTwo}\beta_{\sIdxThree}}{\ulTotPower{0}{}(\ulDuration-1)}
\;;\;
\cValThree
=
\sum\limits_{\sIdxTwo=0}^{N-1}	
\frac{\ulTotPower{\sIdxTwo}{}d_{\sIdxTwo}\ulDuration}{(\ulDuration-1)}
\\		
\cValFour
&=
	\sum\limits_{\sIdxTwo=0}^{N-1}	
	\frac{\ulTotPower{\sIdxTwo}{}e_{\sIdxTwo}\ulDuration}{\ulDuration-1}				
	+
	\frac{\sigma^4}{\ulTotPower{0}{}\ulDuration}
\;;\;
\cValFive
=
\sum\limits_{\sIdxTwo\in\pilotSharingSet_0}
\sigma^2
\frac{\ulTotPower{\sIdxTwo}{}}{\ulTotPower{0}{}}
\beta_{\sIdxTwo}
\;;\;
\cValSix
=
\sum\limits_{\sIdxTwo=0}^{N-1}
\frac{\sigma^2}{\ulDuration}		
\frac{\ulTotPower{\sIdxTwo}{}}{\ulTotPower{0}{}}
\beta_{\sIdxTwo}	
\\		
\cValSeven
&=
\sum\limits_{\pilotSharingSet_{0}\ni\sIdxTwo\neq 0}
\frac{\ulTotPower{\sIdxTwo}{}^2}{\ulTotPower{0}{}}
\beta_{\sIdxTwo}^2
+
\frac{\ulDuration}{\ulDuration-1}
\sum\limits_{\sIdxTwo=0}^{N-1}
\ulTotPower{\sIdxTwo}{}
\lrc{
	\frac{c_{\sIdxTwo}}{\ulDuration}		
	+
	\frac{1}{M}
	d_{\sIdxTwo}
}\;.
\end{align}
Given that $ \signalPower = \expectation\lrf{|s_{\symbolIdx}|^2} = \ulTotPower{0}{}\rhoD{}{}^2M^2\beta_{0}^2 $, substituting \eqref{eqn:interfPowerUpperBoundAlphaSum} into \eqref{eqn:achRateFullFrame}, a lower bound on the UL ergodic capacity is obtained as
\begin{align}
	R_{0}
	=
	\frac{\ulDuration-1}{\ulDuration}
	\logCapacity{\frac{\ulTotPower{0}{}\beta_{0}^2}{			
				\frac{\rhoD{}{}^2}{\rhoP{}{}^2}	
				\cValOne
				+
				\frac{\rhoD{}{}^2}{M\rhoP{}{}^2}	
				\cValTwo
				+
				\frac{
					\rhoP{}{}^2
				}{\rhoD{}{}^2}	
				\frac{\cValThree}{M}
				+	
				\frac{1}{\rhoD{}{}^2\rhoP{}{}^2}
				\frac{\cValFour}{M}
				+
				\frac{1}{\rhoD{}{}^2}
				\frac{\cValFive}{M}
				+
				\frac{1}{\rhoP{}{}^2}
				\frac{\cValSix}{M}
				+
				\cValSeven}}
		\label{eqn:boundAchievableRate}
\end{align}
which is maximized when the denominator inside $ \logCapacity{\cdot} $ is minimized. To obtain $ \rhoD{}{}^2_{\mathrm{opt}} $, we set $ \rhoP{}{}^2 = 1 - \rhoD{}{}^2 $ in \eqref{eqn:boundAchievableRate}, differentiate the denominator with respect to $ \rhoD{}{}^2 $, and set the result to zero. We then get,
\begin{align}
	\rhoD{}{}^2_{\mathrm{opt}}
	=
	\lrc{1 + \sqrt{M} \rhoDrhoPConstant}^{-1}
	\;,\;
	\rhoP{}{}^2_{\mathrm{opt}}
	=
	\lrc{1 + \frac{1}{\rhoDrhoPConstant\sqrt{M}}}^{-1}\;.
\end{align}
where
\begin{equation}
	\rhoDrhoPConstant
	\triangleq
	\sqrt{\frac{\cValOne + \frac{\cValTwo}{M} + \frac{\cValFour}{M}+ \frac{\cValSix}{M}}{\cValThree+\cValFour+\cValFive}}\;.
\end{equation}

%% file: texFiles/appendixMseCrlbFeedback.tex
\section{}
\label{appdx:MSE}
\subsection*{CRLB for Channel Estimates Obtained From SP Pilots}
To derive the CRLB, the received signal when using SP pilots can be written as \cite{upadhya2016superimposed}
\renewcommand{\vIdxOne}{j}
\renewcommand{\vIdxTwo}{m}
\renewcommand{\sIdxOne}{\ell}
\renewcommand{\sIdxTwo}{k}
\renewcommand{\sIdxThree}{k}
\begin{equation}
\label{eqn:superimposedPilotsDefnCrlb}
\mbf{Y} = \mbf{H}_d \lrc{\rhoD{}{}\mbf{X}_d + \rhoP{}{}\mbf{P}_d} + \mbf{H}_i \lrc{\rhoD{}{}\mbf{X}_i + \rhoP{}{} \mbf{P}_i} + \mbf{W}
\end{equation}
where $ \mbf{H}_d \triangleq \lrs{\mbf{h}_{\vIdxOne,\vIdxOne,0},\ldots,\mbf{h}_{\vIdxOne,\vIdxOne,K-1}} $ are the channel vectors of the desired users and $ {\mbf{X}_d \triangleq \lrs{\mbf{x}_{\vIdxOne,0},\ldots,\mbf{x}_{\vIdxOne,K-1}}^T} $ are the data symbols from the desired users. Similarly, $ \mbf{H}_i \in \mathbb{C}^{M\times(\pilotReuseRatioSp-1)K} $ and $ \mbf{X}_i \in \mathbb{C}^{\ulDuration\times(\pilotReuseRatioSp-1)K} $ are the data and channel vectors, respectively, of the interfering users. The subscript $ \vIdxOne $ has been dropped from $ \mbf{H}_d $, $ \mbf{H}_i $, $ \mbf{X}_d $, $ \mbf{X}_i $, and $ \mbf{W} $ for notational convenience. In addition, the UL transmit power $ \ulTotPower{\sIdxOne}{\sIdxTwo} $ for each user is assumed to be absorbed into $ \beta_{\vIdxOne\sIdxOne\sIdxTwo} $. The vectorized form of \eqref{eqn:superimposedPilotsDefnCrlb} can be written as
\renewcommand{\vIdxOne}{j}
\renewcommand{\vIdxTwo}{m}
\renewcommand{\sIdxOne}{\ell}
\renewcommand{\sIdxTwo}{k}
\renewcommand{\sIdxThree}{k}
\begin{align}
\overline{\mbf{y}} = \mathrm{vec}\lrc{\mbf{Y}} &=\lrc{\mbf{I}_{\ulDuration}\otimes\mbf{H}_d} \lrc{\rhoD{}{}\vecmat{x}_d+\rhoP{}{}\vecmat{p}_d} + \lrc{\mbf{I}_{\ulDuration}\otimes\mbf{H}_i} \lrc{\rhoD{}{}\vecmat{x}_i+\rhoP{}{}\vecmat{p}_i} + \vecmat{w} 
\nonumber \\
&= \lrc{\lrc{\rhoD{}{}\mbf{X}_d+\rhoP{}{}\mbf{P}_d}^T \otimes \mbf{I}_M} \vecmat{h}_d + \lrc{\lrc{\rhoD{}{}\mbf{X}_i+\rhoP{}{}\mbf{P}_i}^T \otimes \mbf{I}_M}\vecmat{h}_i + \vecmat{w}
\label{eqn:spRxSignalVec}
\end{align}
where the over-bar denotes the vec operation, i.e., $ \vecmat{x} \triangleq \mathrm{vec}(\mbf{X}) $ and the property $ \mathrm{vec}\lrc{\mbf{A}\mbf{B}} = \lrc{\mbf{I}_m\otimes\mbf{A}}\vecmat{b} = \lrc{\mbf{B}^T\otimes\mbf{I}_n}\vecmat{a} $ has been used. 

For the set of unknown parameters $ \pmb{\theta} \triangleq\lrf{\vecmat{x}_d,\vecmat{x}_i,\vecmat{h}_d,\vecmat{h}_i} $, the Fischer information matrix can be defined as \cite{vandec1994cramer}
\begin{align}
\mathcal{J}\lrc{\pmb{\theta}} &= \mathbb{E}_{\mbf{Y},\pmb{\theta}}\lrf{\lrs{\frac{\partial\;\mathrm{\ln}\;p\lrc{\mbf{Y},\pmb{\theta}}}{\partial\pmb{\theta}^*}}\lrs{\frac{\partial\;\mathrm{\ln}\;p\lrc{\mbf{Y},\pmb{\theta}}}{\partial\pmb{\theta}^*}}^H} 
\nonumber
\\
&= \mathbb{E}_{\pmb{\theta}} \lrf{\mbf{J}_{\pmb{\theta}\pmb{\theta}^H}}+ 
\mathbb{E}_{\pmb{\theta}} \lrf{\lrs{\frac{\partial\mathrm{ln}p\lrc{\pmb{\theta}}}{\partial\pmb{\theta}^*}}\lrs{\frac{\partial\mathrm{ln}p\lrc{\pmb{\theta}}}{\partial\pmb{\theta}^*}}^H} \ . \label{eqn:crlbSimplification}
\end{align}
where
\begin{align}
\mbf{J}_{\pmb{\theta}\pmb{\theta}^H} &\triangleq  \mathbb{E}_{\mbf{Y}\;\vrule\;\pmb{\theta}}\lrf{\lrs{\frac{\partial\mathrm{ln}p\lrc{\mbf{Y}\condProb\pmb{\theta}}}{\partial\pmb{\theta}^*}} \lrs{\frac{\partial\mathrm{ln}p\lrc{\mbf{Y}\condProb\pmb{\theta}}}{\partial\pmb{\theta}^*}}^H \condProb\pmb{\theta}}\;.
\label{eqn:JThetaThetaDefn}
\end{align}
Using \eqref{eqn:spRxSignalVec}, $ \mbf{J}_{\pmb{\theta}\pmb{\theta}^H} $ can be written as
\begin{align}
\mbf{J}_{\pmb{\theta}\pmb{\theta}^H}
& = \begin{bmatrix}
\mbf{J}_{x_d}\mbf{J}_{x_d}^H & \mbf{J}_{x_d}\mbf{J}_{x_i}^H & \mbf{J}_{x_d}\mbf{J}_{h_d}^H & \mbf{J}_{x_d}\mbf{J}_{h_i}^H \\
\mbf{J}_{x_i}\mbf{J}_{x_d}^H & \mbf{J}_{x_i}\mbf{J}_{x_i}^H & \mbf{J}_{x_i}\mbf{J}_{h_d}^H & \mbf{J}_{x_i}\mbf{J}_{h_i}^H \\
\mbf{J}_{h_d}\mbf{J}_{x_d}^H & \mbf{J}_{h_d}\mbf{J}_{x_i}^H & \mbf{J}_{h_d}\mbf{J}_{h_d}^H & \mbf{J}_{h_d}\mbf{J}_{h_i}^H \\
\mbf{J}_{h_i}\mbf{J}_{x_d}^H & \mbf{J}_{h_i}\mbf{J}_{x_i}^H & \mbf{J}_{h_i}\mbf{J}_{h_d}^H & \mbf{J}_{h_i}\mbf{J}_{h_i}^H \\
\end{bmatrix}
\label{eqn:JThetaThetaMatrix}
\end{align}
where
\begin{align}
\label{eqn:JThetaThetaMatrixElementsBegin}
\mbf{J}_{x_d} &\triangleq \frac{\rhoD{}{}}{\sigma} \lrc{\mbf{I}_{\ulDuration}\otimes\mbf{H}_d}^H \;, \quad 
\mbf{J}_{x_i} \triangleq \frac{\rhoD{}{}}{\sigma} \lrc{\mbf{I}_{\ulDuration}\otimes\mbf{H}_i}^H \\
\mbf{J}_{h_d} &\triangleq \frac{1}{\sigma}\lrc{\lrc{\rhoD{}{}\mbf{X}_d+\rhoP{}{}\mbf{P}_d}^T\otimes\mbf{I}_M}^H \;, \quad
\mbf{J}_{h_i} \triangleq \frac{1}{\sigma}\lrc{\lrc{\rhoD{}{}\mbf{X}_i+\rhoP{}{}\mbf{P}_i}^T\otimes\mbf{I}_M}^H 
\ .
\label{eqn:JThetaThetaMatrixElementsEnd}
\end{align}
Using \eqref{eqn:JThetaThetaMatrix} to \eqref{eqn:JThetaThetaMatrixElementsEnd}, the first term in \eqref{eqn:crlbSimplification} can be expressed as
\begin{align}
&\mathbb{E}_{\pmb{\theta}}\lrs{\mbf{J}_{\pmb{\theta}\pmb{\theta}^H}} 
=
\frac{M\rhoD{}{}^2}{\sigma^2}
\mathrm{blkdiag}
\lrs{
\eye_{\ulDuration}\otimes\mbf{D}_d \; , \; {\eye_{\ulDuration}\otimes\mbf{D}_i}\; , \; \frac{\ulDuration}{M\rhoD{}{}^2}\eye_{MK}  \;,\; \frac{\ulDuration}{M\rhoD{}{}^2}\eye_{M(N-K)}}
\end{align}
where $ \mbf{D}_d \triangleq \mathrm{diag}\lrf{\beta_{\vIdxOne,\vIdxOne,0},\ldots,\beta_{\vIdxOne,\vIdxOne,K-1}} $ is the diagonal matrix containing the path-loss coefficients of the desired users  and $ \mbf{D}_i$ is the diagonal matrix containing the path-loss coefficients of the interfering users. The second term in \eqref{eqn:crlbSimplification} can be found as
\begin{align}
&\mathbb{E}_{\pmb{\theta}} \lrf{\lrs{\frac{\partial\mathrm{ln}p\lrc{\pmb{\theta}}}{\partial\pmb{\theta}^*}}\lrs{\frac{\partial\mathrm{ln}p\lrc{\pmb{\theta}}}{\partial\pmb{\theta}^*}}^H} 
=
\mathrm{blkdiag}
\lrs{
	\mbf{\Gamma}_{x_d} \;,\;\mbf{\Gamma_{x_i}}\;,\;\lrc{\eye_M\otimes\mbf{D}_d}^{-1}\;,\;\lrc{\eye_M\otimes\mbf{D}_i}^{-1}
}
\end{align}
where 
\begin{align}
\mbf{\Gamma}_{x_d} &\triangleq \mathbb{E}_{\mbf{x}_d}\lrf{\lrs{\frac{\partial\;\mathrm{ln}\;p_{\mbf{x}_d}\lrc{\mbf{x}_d}}{\partial\mbf{x}_d^*}}\lrs{\frac{\partial\;\mathrm{ln}\;p_{\mbf{x}_d}\lrc{\mbf{x}_d}}{\partial\mbf{x}_d^*}}^H} \\
\mbf{\Gamma}_{x_i} &\triangleq \mathbb{E}_{\mbf{x}_i}\lrf{\lrs{\frac{\partial\;\mathrm{ln}\;p_{\mbf{x}_i}\lrc{\mbf{x}_i}}{\partial\mbf{x}_i^*}}\lrs{\frac{\partial\;\mathrm{ln}\;p_{\mbf{x}_i}\lrc{\mbf{x}_i}}{\partial\mbf{x}_i^*}}^H} \ .
\end{align}
Therefore, the CRLB for the parameter $ \mbf{H}_d $ and the channel vector $ \mbf{h}_{j,j,k} $ are given as
\renewcommand{\vIdxOne}{j}
\renewcommand{\vIdxTwo}{m}
\renewcommand{\sIdxOne}{\ell}
\renewcommand{\sIdxTwo}{k}
\renewcommand{\sIdxThree}{k}
\begin{align}
\mathrm{CRLB}\lrc{\mbf{H}_d} &= \mathrm{trace}\lrf{\lrc{\frac{\ulDuration}{\sigma^2}\eye_{KM}+\lrc{\eye_M\otimes\mbf{D}_d}^{-1}}^{-1}}\\
\mathrm{CRLB}\lrc{\mbf{h}_{\vIdxOne,\vIdxOne,\vIdxTwo}} &= \mathrm{trace}\lrf{\lrc{\frac{\ulDuration}{\sigma^2}\eye_M+\frac{1}{\beta_{\vIdxOne,\vIdxOne,\vIdxTwo}}\eye_M}^{-1}} 
= \frac{M}{\frac{\ulDuration}{\sigma^2}+\frac{1}{\beta_{\vIdxOne,\vIdxOne,\vIdxTwo}}} \ .
\label{eqn:crlbEquation}
\end{align}
This completes the derivation of \eqref{eqn:spCrlb}.

%% file: Main.bbl
\begin{thebibliography}{10}
\providecommand{\url}[1]{#1}
\csname url@samestyle\endcsname
\providecommand{\newblock}{\relax}
\providecommand{\bibinfo}[2]{#2}
\providecommand{\BIBentrySTDinterwordspacing}{\spaceskip=0pt\relax}
\providecommand{\BIBentryALTinterwordstretchfactor}{4}
\providecommand{\BIBentryALTinterwordspacing}{\spaceskip=\fontdimen2\font plus
\BIBentryALTinterwordstretchfactor\fontdimen3\font minus
  \fontdimen4\font\relax}
\providecommand{\BIBforeignlanguage}[2]{{%
\expandafter\ifx\csname l@#1\endcsname\relax
\typeout{** WARNING: IEEEtran.bst: No hyphenation pattern has been}%
\typeout{** loaded for the language `#1'. Using the pattern for}%
\typeout{** the default language instead.}%
\else
\language=\csname l@#1\endcsname
\fi
#2}}
\providecommand{\BIBdecl}{\relax}
\BIBdecl

\bibitem{boccardi2014fivedisruptive}
F.~Boccardi, R.~Heath, A.~Lozano, T.~Marzetta, and P.~Popovski, ``Five
  disruptive technology directions for {5G},'' \emph{{IEEE} Commun. Mag.},
  vol.~52, no.~2, pp. 74--80, Feb. 2014.

\bibitem{jeffandrews2014whatwill}
J.~Andrews, S.~Buzzi, W.~Choi, S.~Hanly, A.~Lozano, A.~Soong, and J.~Zhang,
  ``What will {5G} be?'' \emph{{IEEE} J. Sel. Areas Commun.}, vol.~32, no.~6,
  pp. 1065--1082, Jun. 2014.

\bibitem{jungnickel2014theroleof}
V.~Jungnickel, K.~Manolakis, W.~Zirwas, B.~Panzner, V.~Braun, M.~Lossow,
  M.~Sternad, R.~Apelfröjd, and T.~Svensson, ``The role of small cells,
  coordinated multipoint, and massive {MIMO} in {5G},'' \emph{{IEEE} Commun.
  Mag.}, vol.~52, no.~5, pp. 44--51, May 2014.

\bibitem{larsson2014massive}
E.~Larsson, O.~Edfors, F.~Tufvesson, and T.~Marzetta, ``Massive {MIMO} for next
  generation wireless systems,'' \emph{{IEEE} Commun. Mag.}, vol.~52, no.~2,
  pp. 186--195, Feb. 2014.

\bibitem{osseiran2014scenarios}
A.~Osseiran, F.~Boccardi, V.~Braun, K.~Kusume, P.~Marsch, M.~Maternia,
  O.~Queseth, M.~Schellmann, H.~Schotten, H.~Taoka, H.~Tullberg, M.~A.
  Uusitalo, B.~Timus, and M.~Fallgren, ``Scenarios for {5G} mobile and wireless
  communications: the vision of the {METIS} project,'' \emph{{IEEE} Commun.
  Mag.}, vol.~52, no.~5, pp. 26--35, May 2014.

\bibitem{ngo2013energy}
H.~Q. Ngo, E.~Larsson, and T.~Marzetta, ``Energy and spectral efficiency of
  very large multiuser {MIMO} systems,'' \emph{{IEEE} Trans. Commun.}, vol.~61,
  no.~4, pp. 1436--1449, Apr. 2013.

\bibitem{Marzetta2010Noncooperative}
T.~Marzetta, ``Noncooperative cellular wireless with unlimited numbers of base
  station antennas,'' \emph{{IEEE} Trans. Wireless Commun.}, vol.~9, no.~11,
  pp. 3590--3600, Nov. 2010.

\bibitem{lulu2014anoverview}
L.~Lu, G.~Li, A.~Swindlehurst, A.~Ashikhmin, and R.~Zhang, ``An overview of
  massive {MIMO}: Benefits and challenges,'' \emph{{IEEE} J. Sel. Topics Signal
  Process.}, vol.~8, no.~5, pp. 742--758, Oct. 2014.

\bibitem{Muller2014Blind}
R.~Muller, L.~Cottatellucci, and M.~Vehkapera, ``Blind pilot decontamination,''
  \emph{{IEEE} J. Sel. Topics Signal Process.}, vol.~8, no.~5, pp. 773--786,
  Oct. 2014.

\bibitem{ngo2012evd}
H.~Q. Ngo and E.~Larsson, ``{EVD}-based channel estimation in multicell
  multiuser {MIMO} systems with very large antenna arrays,'' in \emph{Proc.
  IEEE Int. Conf. on Acoustics, Speech and Signal Processing (ICASSP), Kyoto},
  Mar. 2012, pp. 3249--3252.

\bibitem{bjornson2015massive}
E.~Bj\"ornson, E.~G. Larsson, and M.~Debbah, ``Massive {MIMO} for maximal
  spectral efficiency: How many users and pilots should be allocated?''
  \emph{{IEEE} Trans. Wireless Commun.}, vol.~15, no.~2, pp. 1293--1308, Feb.
  2016.

\bibitem{upadhya2015anarray}
K.~Upadhya and S.~A. Vorobyov, ``An array processing approach to pilot
  decontamination for massive {MIMO},'' in \emph{Proc. IEEE 6th Int. Workshop
  on Computational Advances in Multi-Sensor Adaptive Processing (CAMSAP)},
  Cancun, Dec. 2015, pp. 453--456.

\bibitem{Yin2013Coordinated}
H.~Yin, D.~Gesbert, M.~Filippou, and Y.~Liu, ``A coordinated approach to
  channel estimation in large-scale multiple-antenna systems,'' \emph{{IEEE} J.
  Sel. Areas Commun.}, vol.~31, no.~2, pp. 264--273, Feb. 2013.

\bibitem{bjornson2017pilot}
E.~Bj\"{o}rnson, J.~Hoydis, and L.~Sanguinetti, ``Pilot contamination is not a
  fundamental asymptotic limitation in massive mimo,'' in \emph{IEEE
  International Conference on Communications (ICC)}, May 2017, pp. 1--6.

\bibitem{takeuchi2013Achievable}
K.~Takeuchi, R.~R. M\"{u}ller, M.~Vehkaper\"{a}, and T.~Tanaka, ``On an
  achievable rate of large rayleigh block-fading {MIMO} channels with no
  {CSI},'' \emph{{IEEE} Trans. Inf. Theory}, vol.~59, no.~10, pp. 6517--6541,
  Oct. 2013.

\bibitem{shuangchi2007superimposed}
S.~He, J.~Tugnait, and X.~Meng, ``On superimposed training for {MIMO} channel
  estimation and symbol detection,'' \emph{{IEEE} Trans. Signal Process.},
  vol.~55, no.~6, pp. 3007--3021, Jun. 2007.

\bibitem{coldrey2007training}
M.~Coldrey and P.~Bohlin, ``Training-based {MIMO} systems -- part {I}:
  Performance comparison,'' \emph{{IEEE} Trans. Signal Process.}, vol.~55,
  no.~11, pp. 5464--5476, Nov. 2007.

\bibitem{haidong2003pilot}
H.~Zhu, B.~Farhang-Boroujeny, and C.~Schlegel, ``Pilot embedding for joint
  channel estimation and data detection in {MIMO} communication systems,''
  \emph{{IEEE} Commun. Lett.}, vol.~7, no.~1, pp. 30--32, Jan. 2003.

\bibitem{cui2005pilot}
T.~Cui and C.~Tellambura, ``Pilot symbols for channel estimation in {OFDM}
  systems,'' in \emph{Proc. IEEE Global Telecommunications Conf. (GLOBECOM),
  St. Louis}, vol.~4, Dec. 2005, pp. 5 pp.--2233.

\bibitem{upadhya2016superimposed}
K.~Upadhya, S.~A. Vorobyov, and M.~Vehkapera, ``Superimposed pilots are
  superior for mitigating pilot contamination in massive {MIMO},'' \emph{{IEEE}
  Trans. Signal Process.}, vol.~65, no.~11, pp. 2917--2932, Jun. 2017.

\bibitem{verenzuela2017spectral}
\BIBentryALTinterwordspacing
D.~Verenzuela, E.~Bj{\"{o}}rnson, and L.~Sanguinetti, ``Spectral and energy
  efficiency of superimposed pilots in uplink massive {MIMO},'' \emph{arXiv
  Preprint}, 2017. [Online]. Available: \url{http://arxiv.org/abs/1709.07722}
\BIBentrySTDinterwordspacing

\bibitem{Zhang2015SuperimposedPilot}
H.~Zhang, S.~Gao, D.~Li, H.~Chen, and L.~Yang, ``On superimposed pilot for
  channel estimation in multi-cell multiuser {MIMO} uplink: Large system
  analysis,'' \emph{{IEEE} Trans. Veh. Technol.}, vol.~65, no.~3, pp.
  1492--1505, Mar. 2016.

\bibitem{upadhya2016downlink}
K.~Upadhya, S.~A. Vorobyov, and M.~Vehkapera, ``Downlink performance of
  superimposed pilots in massive {MIMO} systems in the presence of pilot
  contamination,'' in \emph{Proc. IEEE Global Conf. on Signal and Information
  Processing (GlobalSIP)}, Washington D.C., Dec. 2016, pp. 665--669.

\bibitem{upadhya2016hybrid}
------, ``Time-multiplexed / superimposed pilot selection for massive {MIMO}
  pilot decontamination,'' in \emph{Proc. IEEE Int. Conf. on Acoustics, Speech
  and Signal Processing (ICASSP)}, New Orleans, LO, Mar. 2017, pp. 3459 --
  3463.

\bibitem{hoydis2013massive}
J.~Hoydis, S.~ten Brink, and M.~Debbah, ``Massive {MIMO} in the {UL/DL} of
  cellular networks: How many antennas do we need?'' \emph{{IEEE} J. Sel. Areas
  Commun.}, vol.~31, no.~2, pp. 160--171, Feb. 2013.

\bibitem{jose2011pilot}
J.~Jose, A.~Ashikhmin, T.~Marzetta, and S.~Vishwanath, ``Pilot contamination
  and precoding in multi-cell {TDD} systems,'' \emph{{IEEE} Trans. Wireless
  Commun.}, vol.~10, no.~8, pp. 2640--2651, Aug. 2011.

\bibitem{nayebi2018semi-blind}
E.~Nayebi and B.~D. Rao, ``Semi-blind channel estimation for multiuser massive
  {MIMO} systems,'' \emph{{IEEE} Trans. Signal Process.}, vol.~66, no.~2, pp.
  540--553, Jan 2018.

\bibitem{Hassibi2003HowMuch}
B.~Hassibi and B.~M. Hochwald, ``How much training is needed in
  multiple-antenna wireless links?'' \emph{{IEEE} Trans. Inf. Theory}, vol.~49,
  no.~4, pp. 951--963, Apr. 2003.

\bibitem{kong2016Channel}
D.~Kong, D.~Qu, K.~Luo, and T.~Jiang, ``Channel estimation under staggered
  frame structure for massive {MIMO} system,'' \emph{{IEEE} Trans. Wireless
  Commun.}, vol.~15, no.~2, pp. 1469--1479, Feb. 2016.

\bibitem{mahyiddin2015performance}
W.~A. W.~M. Mahyiddin, P.~A. Martin, and P.~J. Smith, ``Performance of
  synchronized and unsynchronized pilots in finite massive {MIMO} systems,''
  \emph{{IEEE} Trans. Wireless Commun.}, vol.~14, no.~12, pp. 6763--6776, Dec.
  2015.

\bibitem{vandec1994cramer}
A.~Van Den~Bos, ``A {C}ram{\'e}r-{R}ao lower bound for complex parameters,''
  \emph{{IEEE} Trans. Signal Process.}, vol.~42, no.~10, pp. 2859--, Oct. 1994.

\end{thebibliography}
